\documentclass[12pt,a4paper]{elsarticle}

\usepackage{graphicx}
\usepackage{multirow}
\usepackage{xcolor}

\usepackage{lineno,hyperref}
\modulolinenumbers[5]
\journal{Applied Mathematical Modelling}

\biboptions{sort&compress}

\usepackage{amsmath,amssymb,amsthm,amsfonts}
\usepackage{nccmath}
\usepackage{mathtools}
\usepackage{epsfig}
\usepackage{siunitx}	
\usepackage{caption}
\usepackage{subcaption}
\usepackage{float}
\usepackage{verbatim}
\usepackage[font=small]{caption}
\usepackage{soul}
\usepackage{multirow}
\usepackage{url}
\usepackage[utf8]{inputenc}
\usepackage[T1]{fontenc}
\usepackage{color}
\usepackage{booktabs,overpic}

\usepackage{tikz}

\usepackage{physics}
\usepackage{enumerate}

\def\k{\kappa}

\def\p{\partial}

\def\({\text{\huge (}}
\def\){\text{\huge )}}

\def\]{\text{\huge ]}}
\def\[{\text{\huge [}}

\newcommand\nc{\newcommand}

\nc{\red}[1]{\textcolor{red}{#1}}
\nc{\Oc}{\mathcal{O}}
\nc{\refb}[1] {(\ref{#1})}		 	   
\nc{\fig}[1] {figure \ref{#1}}	   
\nc{\Fig}[1] {Figure \ref{#1}}	   
\nc{\figs}[1] {figures \ref{#1}}	   
\nc{\Figs}[1] {Figures \ref{#1}}	   
\nc{\sect}[1] {Section \ref{#1}}       
\nc{\Sect}[1] {Section \ref{#1}}       
\nc{\chap}[1] {chapter \ref{#1}}       
\nc{\Chap}[1] {Chapter \ref{#1}}       
\nc{\app}[1] {\ref{#1}}       
\nc{\App}[1] {\ref{#1}}       
\nc{\ud}{\mathop{}\!\mathrm{d}}
\nc{\uD}{\mathop{}\!\mathrm{D}}
\nc{\bcdot}{\boldsymbol{\cdot}}
\nc{\lab}[1]{\label{#1} \quad #1\quad} 
\nc{\et}{\emph{et al}.\ }
\nc{\bi}{\begin{itemize}}
\nc{\ei}{\end{itemize}}
\nc{\be}{\begin{equation}}
\nc{\ee}{\end{equation}}
\nc{\ba}{\begin{align}}
\nc{\ea}{\end{align}}
\nc{\non}{\nonumber\\}

\nc\pad[2]{\frac{\p #1}{\p #2}} 
\nc\padd[2]{\frac{\p^2 #1}{\p
{#2}^2}} 
\nc\nd[2]{\frac{\mathrm{d} #1}{\mathrm{d} #2}} 
\nc\ndd[2]{\frac{\mathrm{d}^2 #1}{\mathrm{d}
{#2}^2}} 
\nc\pat[2]{\frac{\mathrm{D} #1}{\mathrm{D}
#2}} \nc\ov{\overline} 
\nc\degree{^{\circ}} 
\nc\ord[1]{{\cal
O}(#1)} 
\nc\ra{\rightarrow} 
\nc\Ra{\Rightarrow} 
\nc\dint{{\mbox ~
d}}
\nc{\units}[1]{$^\mathrm{#1}$}
\nc\T{\hat{t}}
\nc\C{\hat{c}}
\nc\X{\hat{x}}
\nc\Q{\hat{q}}
\nc\V{\hat{v}}
\nc\RI{\mbox{R}_{I}}
\nc\RII{\mbox{R}_{II}}

\DeclareMathOperator{\Pe}{Pe}       
\DeclareMathOperator{\Da}{Da}		

\nc{\bea}{\begin{eqnarray}}
\nc{\eea}{\end{eqnarray}}
\nc{\beas}{\begin{eqnarray*}}
\nc{\eeas}{\end{eqnarray*}}

\nc{\abel}[1]{\textcolor{red}{#1}}
\nc{\marc}[1]{\textcolor{magenta}{#1}}
\nc{\tim}[1]{\textcolor{blue}{#1}}
\nc{\raj}[1]{\textcolor{purple}{#1}}
\nc{\fran}[1]{\textcolor{green}{#1}}
\nc\hc{ \hat{c}}   
\nc\hq{ \hat{q}}

\date{\today}

\begin{document}

\begin{frontmatter}
\title{On the use of equilibrium models to describe dynamic adsorption regimes}

\author[upc2,crm,upc3]{M. Calvo-Schwarzwalder}
\author[upc1]{A. Valverde\fnref{myfootnote}}
\author[upc4]{A. Cuesta López}
\author[lequia]{A. Cabrera-Codony}
\author[crm]{U. Thorat}
\author[crm]{T.G. Myers}

\address[upc2]{Department of Mathematics, Universitat Polit\`ecnica de Catalunya, Escola d'Enginyeria de Telecomunicació i Aeroespacial de Castelldefels, 08860 Castelldefels, Spain}
\address[crm]{Centre de Recerca Matem\`atica, Campus de Bellaterra, Edifici C, 08193 Bellaterra, Barcelona, Spain}
\address[upc3]{Institut de Matem\`atiques-BarcelonaTech, Universitat Polit\`ecnica de Catalunya, Facultat de Matem\`atiques i Estadistica, 08860 Barcelona, Spain}
\address[upc1]{Department of Chemical Engineering, Universitat Polit\`ecnica de Catalunya, Escola Superior d’Enginyeries Industrial, Aeroespacial i Audiovisual de Terrassa, 08222 Terrassa, Spain}
\address[upc4]{Universitat Polit\`ecnica de Catalunya, Facultat de Matem\`atiques i Estadistica, 08860 Barcelona, Spain}
\address[lequia]{LEQUIA, Institute of the Environment, Universitat de Girona, 17003 Girona, Catalonia, Spain}

\fntext[myfootnote]{Corresponding author: abel.valverde@upc.edu}
\begin{abstract}
    We present a column adsorption model that couples a Pseudo-First-Order (PFO) kinetic formulation with the Sips isotherm framework. Using a traveling wave approximation, we derive analytical solutions for specific operating conditions. Qualitatively, these solutions deviate significantly from their pure Sips counterparts: instead of a smooth, continuous increase in concentration at the column outlet, the PFO-Sips model predicts an abrupt, sudden breakthrough. We validate these analytical solutions against diverse experimental datasets from the literature. The results reveal that the PFO-based model consistently underperforms compared to the original Sips formulation. Furthermore, this validation exposes fundamental inconsistencies within the PFO framework. We demonstrate that despite its widespread use in the literature for almost a century, the PFO model is inherently flawed and structurally unfit for describing column adsorption dynamics.
\end{abstract}


\end{frontmatter}

\section{Introduction}
Adsorption in packed columns is a widely used separation technology, with
applications spanning the removal of volatile organic compounds from industrial
off-gases \cite{Tefera2014,Tzanakopoulou2024}, the capture of CO$_2$ from
flue gas \cite{Casas2012,Myers20a,Liu2022} and the treatment of
contaminated water \cite{NAIDU2021117955,Sulaymon2009}. The
mathematical description of the process requires coupling a mass transfer
equation, governing the transport of the contaminant along the column, with a
kinetic equation, describing the rate at which the contaminant is captured by the
adsorbent \cite{Ruthven1984}. A fundamental requirement of any well-posed model
is internal consistency between these two components: the steady state of the
kinetic equation must yield the adsorption isotherm, that is, the equilibrium
relationship between the fluid-phase concentration and the adsorbed amount.
This requirement is not merely formal. The isotherm determines the equilibrium
capacity of the adsorbent and, through this, the propagation speed of the
adsorption front. A kinetic equation that does not recover the correct isotherm at
steady state will predict the wrong front speed and the wrong breakthrough
behaviour, regardless of how well its parameters are fitted to a particular dataset.
The Sips kinetic model \cite{Sips48} satisfies this consistency requirement. It
describes adsorption and desorption through a mass-action law involving
stoichiometric coefficients $m$ and $n$, whose steady state yields the Sips
isotherm with a characteristic exponent $\alpha = m/n$. For $m = n = 1$ the
model reduces to Langmuir kinetics \cite{Langmuir1918}. Analytical travelling wave
solutions for the consistent Sips model, valid for integer $m, n \geq 1$, have been
derived in \cite{Aguareles2022,SensiTW}, providing closed-form
breakthrough curves whose parameters are directly related to the adsorption and
desorption rate constants. The Sips isotherm is functionally identical to the Hill equation, introduced to describe the cooperative binding of oxygen to
haemoglobin \cite{Hill1910} and now ubiquitous in pharmacology and biochemistry
as a model for dose-response curves and cooperative ligand binding \cite{Goutelle2008}. The mathematical equivalence of the Hill and Sips isotherms has been
formally established by Chu et al. \cite{Chu2022}. The mathematical results derived here
therefore apply directly to any system governed by Hill-type kinetics.

Despite the availability of consistent kinetic formulations, one of the most widely
used approximations in column adsorption modelling is the pseudo-first order (PFO) or Lagergren model \cite{Lage98}. In its original form, this model states that the rate of adsorption is proportional to the difference between the equilibrium amount adsorbed and the current amount. However, the model provides no functional relationship between the mentioned equilibrium value and the fluid-phase concentration, and therefore defines no isotherm. 
To restore the missing concentration dependence, the standard practice involves two modifications. First, the equilibrium value $q_\text{e}$ is defined using the isotherm from a different kinetic model, most commonly Langmuir or Sips. Second, the inlet concentration in this borrowed isotherm is replaced by the local, spatially varying concentration, so that the equilibrium quantity becomes a field variable. The resulting formulation is neither first order, due to the typically non-linear dependence on concentration, nor does it describe an equilibrium state, since this quantity now varies with position and time. Nevertheless, the equilibrium origin of the expression is routinely obscured by referring to it  as the \emph{``equilibrium relationship''} \cite{NAIDU2021117955}. 
When the borrowed isotherm belongs to the Langmuir model, the PFO may also be called the Langmuir Linear Driving Force Model \cite{Fenti2018_LLDF}. Numerous numerical studies employ this approach \cite{Casas2012,Tefera2014,NAIDU2021117955,Tzanakopoulou2024} and comprehensive reviews present it as the standard modelling framework for fixed-bed adsorption \cite{Shafeeyan2014}. We note that the same type of inconsistency arises when the pseudo-second order model \cite{Ho1999} is combined with a borrowed isotherm, though we restrict the present analysis to
the PFO case.

The origin of this approximation can be traced to Thomas \cite{Thomas1944}, who introduced the substitution of the local concentration into the equilibrium isotherm to permit analytical progress. Thomas himself described the step as \emph{``mathematically arbitrary''}. In an era without numerical methods capable of solving the coupled,
nonlinear system, this pragmatic choice was understandable. Today, however, neither justification survives: modern solvers handle the consistent formulation without difficulty, and analytical travelling wave solutions are available for both
Langmuir and Sips kinetics \cite{Myers23,Aguareles2022,SensiTW}. A parallel critique of the Bohart-Adams model, another widely used inconsistent formulation, has been presented by Myers \cite{Myers24}. Yet the PFO approach persists, for at least two reasons. First, it is simpler to implement: the PFO introduces a single kinetic parameter, $k_\text{LDF}$, in place of the two rate constants $k_a$ and $k_d$ of the Sips model, and the nonlinearity enters only through the isotherm expression rather than through the full mass-action law.
Second, the approach has acquired the weight of convention. Decades of published studies employing the PFO formulation, using either Sips or Langmuir as the underlying models, create an inertia that is difficult to overcome, particularly when the fitted curves appear to describe experimental data reasonably well. This last point merits emphasis. A good fit to a single breakthrough experiment does not validate a model; it merely confirms that the model has enough flexibility to reproduce one dataset. The critical test is whether the extracted parameters have physical meaning and whether the model is predictive under conditions different from those used for calibration: a different inlet concentration, a different bed length, or a different adsorbent loading. A model with an inconsistent kinetic-isotherm pairing may absorb the inconsistency into effective parameter values that change unpredictably from one experiment to the next, producing good individual fits but no transferable understanding. This distinction between curve fitting and physical modelling is at the core of the present work.

Given the continued and widespread use of the PFO formulation, the purpose of this paper is to subject it to a rigorous mathematical analysis within the travelling wave framework and to assess whether it can serve as a reliable surrogate for the consistent Sips model. First, we establish the conditions under which the PFO model admits travelling wave solutions. We show that such solutions exist if and only if a $\alpha\leq 1$, where $\alpha$ is the characteristic exponent from the Sips isotherm. Second, for a few specific cases of interest ($\alpha = 1, 1/2, 1/3$ and $2/3$) we derive implicit analytical expressions for the breakthrough curve. 
These reveal a qualitative difference with the consistent Sips model: for $\alpha < 1$ the breakthrough is abrupt, with a finite time $t_b$ before which the outlet concentration is identically zero, in contrast to the smooth onset predicted by the Sips kinetics. This means that the two models are not merely quantitatively different but structurally distinct, producing breakthrough curves of different functional form. Third, we fit both models to experimental
breakthrough data from three independent studies involving different
contaminants and adsorbents \cite{Chuang2003,Sulaymon2009,Liu2022} and compare the resulting parameters and quality of fit, to determine whether the PFO formulation retains any physical meaning.

The paper is organized as follows. In Section~\ref{sec:2} we introduce the mathematical formulation and present both Sips and PFO models, which are then scaled and reduced via a non-dimensionalisation process in Section~\ref{sec:3}. The travelling wave solutions are derived in Section~\ref{sec:4} and validated against numerical solutions and experimental data in Section~\ref{sec:5}. 
The article concludes in Section~\ref{sec:6}.

\section{Mathematical formulation}\label{sec:2}

\subsection{Problem geometry, operational conditions and key variables}
We consider a column of length $L$ and with a circular cross-section of radius $R$. The column is filled with a porous material, termed the adsorbent, which occupies a fraction $1-\epsilon$, with $\epsilon$ is called the cross-sectional void fraction and is assumed constant. 

\begin{figure}
    \centering
    \includegraphics[width=0.75\linewidth]{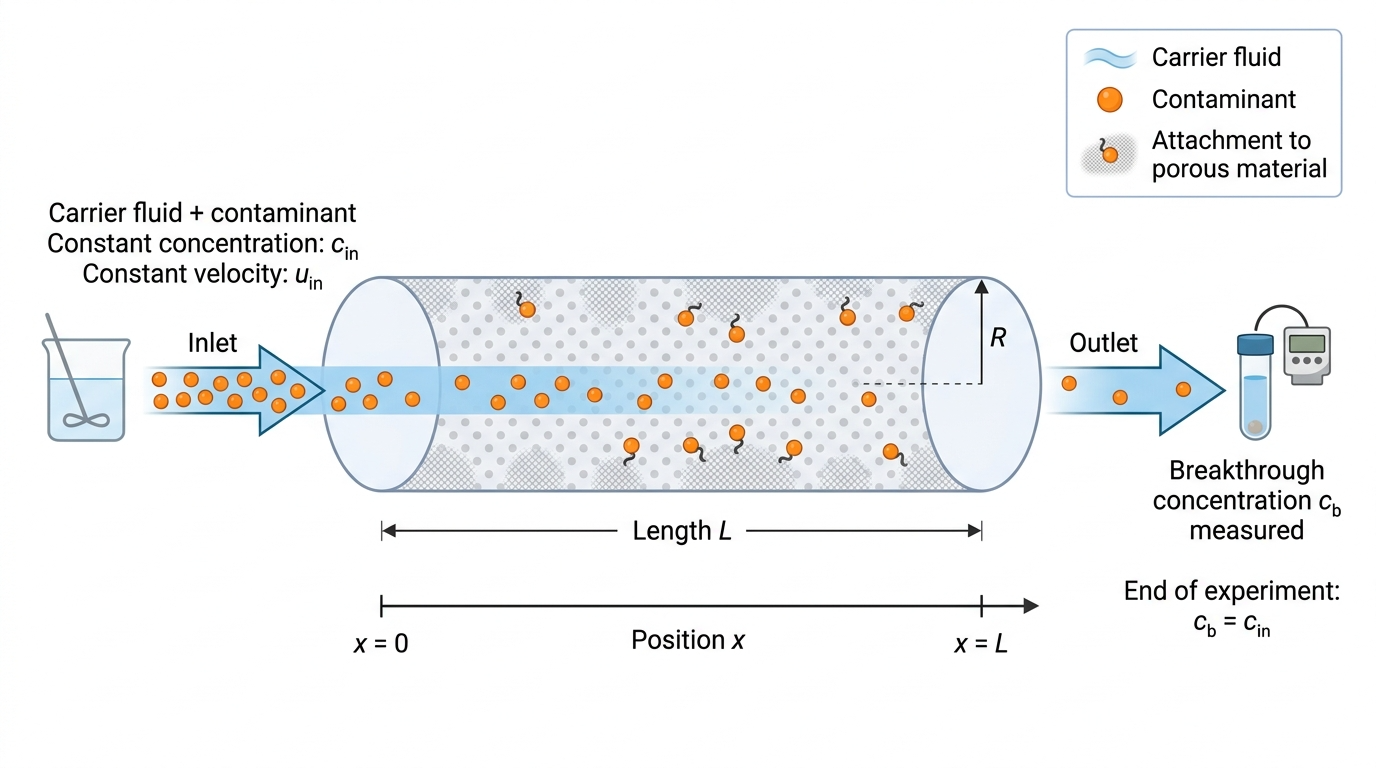}
    \caption{Schematic of the experimental setup.}
    \label{fig:scheme}
\end{figure}

Once the experiment begins, a carrier gas containing a  constant concentration $c_\text{in}$ of the species to be removed, termed adsorbate, is introduced into the column through the inlet ($x=0$), at a constant velocity $u_\text{in}$. For this work, we consider only trace amounts to be adsorbed, which causes the velocity $u$ to be constant along the entire column \cite{Myers20a,Myer22,Myers23}, $u\equiv u_\text{in}$. In the fluid, $c=c(x,t)$ represents the molar concentration of the adsorbate, while $q=q(x,t)$ describes the adsorbed amount per unit adsorbent. Whilst being adsorbed, the mixture flows towards the outlet ($x=L$) to exit the column, where the breakthrough concentration $c_b(t):=c(L,t)$ is measured until the end of the experiment. A schematic of this is provided in Figure~\ref{fig:scheme}. The process is assumed to be isothermal and approximately isobaric, as suggested by the slow flow. 

\subsection{Mass transfer equation}
The mathematical description of the process consists of two equations. The mass transfer equation describes the flow of the contaminant along the column and how the concentration of the contaminant evolves due to advection, dispersion and adsorption. For trace amounts, this equation takes the form
\begin{equation}\label{eq:masstransfer}
    \pad{c}{t}+u_\text{in}\pad{c}{x}=D\padd{c}{x}-\frac{\rho_b}{\epsilon}\pad{q}{t}\, .
\end{equation}
where $D$ is the dispersion coefficient, $\rho_b$ is the bulk density. 

\subsection{Kinetic model}
The second equation of the model is the one describing how the adsorption process occurs. Throughout the literature, many authors have used a large variety of kinetic models, see for instance a review in \cite{Shafeeyan2014}. We will split the description into the two main regimes: the dynamic regime, where $c$ and $q$ are variable, and the equilibrium regime, where both $c$ and $q$ have reached an equilibrium value and adsorption no longer occurs.

\subsubsection{Dynamic regime}
In general, the kinetic model may be written down as 
\begin{equation}\label{eq:kinetic}
    \pad{q}{t}=F(c,q)\, .
\end{equation}

The basis description for the kinetics in this work is the Sips model \cite{Sips48},
\begin{equation}\label{eq:sips}
    F(c,q)=k_ac^m\left(q_\text{max}-q\right)^n-k_dq^n\, ,
\end{equation}
where $n,m$ are stochiometric coefficients determined by the chemical reaction, and $k_a$ and $k_d$ are the adsorption and desorption rates. For $n=m=1$, Eq.~\eqref{eq:sips} reduces to the Langmuir kinetic model \cite{Langmuir1918}. Whereas the prior is typically associated to chemical adsorption via the coefficients $m,n$, the latter is related to physical adsorption.

\subsubsection{Equilibrium state}
At equilibrium we have
\begin{equation}
    c\to c_\text{in},\quad q\to q_\text{e}\, \quad \pad{q}{t}\to0\, ,
\end{equation}
hence $F\to0$, which leads to
\begin{equation}
    F(c_\text{in},q_\text{e})=0\, ,
\end{equation}
and from where we obtain the relation between concentration and adsorbed amount at equilibrium, $q_\text{e}=q_\text{e}(c_\text{in})$, known as the isotherm. 

Note, the isotherm depends on the underlying kinetic model, as it depends on the form of $F$ being used. For the Sips kinetic equation, Eq.~\eqref{eq:sips}, the isotherm
\begin{equation}\label{eq:isosips}
    q_\text{e}(c_\text{in})=\frac{q_\text{max}k_a^{1/n}c_\text{in}^{m/n}}{k_d^{1/n}+k_a^{1/n}c_\text{in}^{m/n}}=:\frac{q_\text{max}K_Sc_\text{in}^{m/n}}{1+K_Sc_\text{in}^{m/n}},
\end{equation}
where we have introduced the Sips equilibrium constant
\begin{equation}\label{eq:Kskd}
    K_S=\left(\frac{k_a}{k_d}\right)^{1/n}\, .
\end{equation}
Setting $m=n=1$ leads to the Langmuir isotherm.

\subsubsection{Pseudo-first order model}
In the literature it is common to find dynamic formulations based on isotherm models. The simplest example is the linear driving force (LDF) model, which reads
\begin{equation}\label{eq:LDFstatic}
    F(c,q)=k_{LDF}\left(q_\text{e}^{*}-q\right)\, ,
\end{equation}
where $q_\text{e}^{*}=q_\text{e}(c_\text{in})$. Equation~\eqref{eq:LDFstatic} states that the adsorbed amount increased along the column at a rate $k_{LDF}$ and independent of whether contaminant is present or not at any specific point $x$. In particular, $k_{LDF}$ is independent of the concentration entering the column, too. This formulation is clearly unphysical and has already been discussed in \cite{Myers20a,Myers23,Valverde2024}. In particular, in \cite{Valverde2024} it is shown that this model can lead to negative values for the concentration. 

Besides de static LDF model, it is not unusual to find a dynamic LDF model, where the equilibrium value $c_\text{in}$ in the equilibrium value $q_\text{e}$ isotherm is substituted by the variable $c$, leading to   
\begin{equation}\label{eq:LDFdynamic}
    F(c,q)=k_{LDF}\left(q_\text{e}(c)-q\right)\, .
\end{equation}
This is termed the pseudo-first order (PFO) model, as it resembles to a first order differential equation for $q$. However, the (typically) non-linear term $q_\text{e}(c)$ introduces a much higher complexity than Eq.~\eqref{eq:LDFstatic}. In particular, in areas where no contaminant is present we have $q_\text{e}(0)=0$ and hence no adsorption occurs.

In the literature, the terms linear driving force and pseudo-first order are sometimes assumed equivalent. Here we clearly separate the names as Eq.~\eqref{eq:LDFstatic} is the only equation that is truly linear.

\subsection{Boundary and initial conditions}
The system is closed with the boundary conditions
\begin{subequations}\label{eq:bcic}
\bea
\label{eq:bc}
u_\text{in} c_\text{in}=\left(u_\text{in} c-D\pad{c}{x}\right)\Bigg\vert_{x=0^+}  ,\qquad \pad{c}{x}\bigg|_{x=L^-} = 0\,,\quad \textrm{for $t>0$} ,
\eea
and the initial conditions
\bea\label{eq:ic}
c(x,0)=q(x,0)=0\, ,\quad \textrm{for $x\in(0,L)$}.
\eea
\end{subequations}

\subsection{Summary of the governing equations}
The final model consists of Eqs.~\eqref{eq:masstransfer} and \eqref{eq:kinetic}, which are subject to the boundary and initial conditions provided in Eq.~\eqref{eq:bcic}. The most important part is the choice of kinetic model $F(c,q)$, as this described the adsorption occurs. The aim of this article is to compare the Sips kinetic model, Eq.~\eqref{eq:sips}, to the kinetic model obtained by using the Sips isotherm, Eq.~\eqref{eq:isosips}, in combination with the PFO model described in Eq.~\eqref{eq:LDFdynamic}. When the latter is being used, we will refer to it as the Sips-PFO (S-PFO) model. Note, for $m=n=1$ the model reduces to considering the Langmuir kinetic or the Langmuir-PFO (L-PFO) models.

\section{Non-dimensionalisation}\label{sec:3}

\subsection{Scaled equations}
We introduce the scaled quantities
\begin{equation}
    \hat{c}=\frac{c}{c_\text{in}},\quad \hat{q}=\frac{q}{q_\text{max}},\quad \hat{x}=\frac{x}{\ell},\quad \hat{t}=\frac{t}{\tau},
\end{equation}
and Eq.~\eqref{eq:masstransfer} becomes
\begin{equation}\label{nd:masstransfer}
    \Da\pad{\hat c}{\hat t}+\pad{\hat c}{\hat x}=\Pe^{-1}\padd{\hat c}{\hat x}-\frac{\rho_b \ell q_\text{max}}{u\epsilon c_\text{in}\tau}\pad{\hat q}{\hat t}\, .
\end{equation}
with
\begin{equation}
    \Da=\frac{\ell}{u\tau},\qquad \Pe^{-1}=\frac{D}{\ell u}.
\end{equation}
To balance advection with adsorption, we choose $\ell={u\epsilon c_\text{in}\tau}/(\rho_b q_\text{max})$. 

As for $\tau$, we will use the value that balances the time derivative and the adsorption term in Eq.~\eqref{eq:sips}. In the non-dimensional formulation, the dynamic Sips model becomes
\begin{equation}\label{nd:sipsPre}
    \frac{1}{\tau}\pad{\hat q}{\hat t}=k_ac_\text{in}^mq_\text{max}^{n-1}\hat c^m\left(1-\hat q\right)^n-k_dq_\text{max}^{n-1}\hat q^n\, ,
\end{equation}
hence we balance the time derivative with the adsorption term setting $\tau=1/(k_ac_\text{in}^mq_\text{max}^{n-1})$. The final form of Eq.~\eqref{nd:sipsPre} is 
\begin{equation}
    \label{nd:sips}
    \pad{\hat q}{\hat t}=\hat c^m\left(1-\hat q\right)^n-\frac{k_d}{k_ac_\text{in}^m}\hat q^n=:\hat c^m\left(1-\hat q\right)^n-\kappa\hat q^n\, .
\end{equation}
The relation between $\kappa$ and the equilibrium constant $K_S$ is $\kappa=K_S^{-n}c_\text{in}^{-m}$. Using these scaled quantities, the Sips isotherm becomes
\begin{equation}
    \frac{q_\text{e}^{*}}{q_\text{max}}=\frac{1}{1+\kappa^{1/n}}=:\hat q_\text{e}^{*}.
\end{equation}
The non-dimensional S-PFO model reads
\begin{equation}
    \label{nd:pfo}
    \frac{1}{\gamma}\pad{\hat q}{\hat t}=\frac{\hat c^{\alpha}}{\kappa^{1/n} + \hat c^{\alpha}}-\hat q\, ,
\end{equation}
with 
\begin{equation}
    \alpha=\frac{m}{n},\qquad \gamma=\frac{k_{LDF}}{k_ac_\text{in}^mq_\text{max}^{n-1}}.
\end{equation}
Finally, the non-dimensional boundary and initial conditions are
\begin{subequations}\label{nd:bcic}
\bea
\label{nd:bc}
1=\left(\hat c-\Pe^{-1}\pad{\hat c}{\hat x}\right)\Bigg\vert_{\hat x=0^+}  ,\qquad \pad{\hat c}{\hat x}\bigg|_{\hat x=\hat L^-} = 0\,,\quad \textrm{for $\hat t>0$} ,
\eea
and the initial conditions
\bea\label{nd:ic}
\hat c(\hat x,0)=\hat q(\hat x,0)=0\, ,\quad \textrm{for $\hat x\in(0,\hat L)$},
\eea
\end{subequations}
where $\hat L=L/\ell$. 

\subsection{S-PFO model without desorption}
In the limit $\kappa\to0$, which occurs as we assume $k_ac_\text{in}\gg k_d$ and hence implies that desorption becomes negligible in the process, we note that Eq.~\eqref{nd:pfo} reduces to
\begin{equation}\label{nd:pfo no k}
    \frac{1}{\gamma}\pad{\Q}{\T}=1-\Q.
\end{equation}
In particular, the latter is independent of the stochiometric parameters $m$ and $n$. In fact, Eq.~\eqref{nd:pfo no k} is the non-dimensional form of the LDF model provided in Eq.~\eqref{eq:LDFstatic} with $q_\text{e}=q_\text{max}$.

\subsection{Parameter values}
The Damköhler and inverse Péclet numbers, $\Da$ and $\Pe^{-1}$, are typically considered small. For the first, this is based on the assumption that the reaction time scale is much larger than the flow time scale, since the process can last up to the order of hours, whereas the flow reaches the outlet in matter of seconds or minutes. For the second, the claim is based on the fact that the process is much more advection-driven than dispersion-driven. 

The parameter $\kappa$ describes the capacity of the adsorbent to retain the contaminant it has already removed from the carrier fluid. As its value decreases, desorption occurs less frequently, and in the limit as $\kappa\to0$ it is neglected completely. Therefore, this parameter is expected to be small.

For the parameter $\gamma$ we will not make any particular assumption, we will simply suppose that $\gamma=\ord{1}$, which may be explained by the fact that the real adsorption time scale should be independent of the kinetic model being used. However, as $k_{LDF}$ does not describe adsorption only, but rather how fast $q$ approaches $q_\text{e}$, in general we assume $k_{LDF}\neq k_ac_\text{in}^mq_\text{max}^{n-1}$, i.e., $\gamma\neq1$.

\subsection{Model reduction}
Upon assuming $\Da,\Pe^{-1}\ll1$, we neglect the corresponding terms in Eq.~\eqref{nd:masstransfer} and obtain
\begin{equation}\label{red:masstransfer}
    \pad{\hat c}{\hat x}=-\pad{\hat q}{\hat t}\, .
\end{equation}
The boundary condition at the inlet becomes
\bea
\label{nd:bc2}
\hat c(0^+,\hat t)=1,
\eea
stating that we have passed an initial transient time where the concentration at the inlet increases due to the influx, but adsorption has not yet begun (since we do not need the initial condition for $\hat c$, but we do need the one for $\hat q$ still).

\section{Approximate solutions in the form of travelling waves}\label{sec:4}
\subsection{Travelling wave formulation}
As done in previous studies, we will aim to solve the equations using a travelling wave approach. For this we introduce $\eta=\hat x-\hat L-\hat v(\hat t-\hat t_h)$ and assume
\begin{equation}
    \hat c\left(\hat x,\hat t\right)=:\mathcal{C}(\eta)\, ,\qquad \hat q\left(\hat x,\hat t\right)=:\mathcal{Q}(\eta)
\end{equation}
The (scaled) time $\hat t_h$ is called the half-time and represents the instant when half of the inlet concentration is measured at the outlet, i.e., $\hat c(\hat L,\hat t_h)=1/2$. In defining $\eta$ in this way, $\eta=0$ corresponds to the position in the wave where $\mathcal{C}=1/2$.

Using these new variables in Eq.~\eqref{red:masstransfer} we find
\begin{equation}\label{TW:relation}
    \nd{\mathcal{C}}{\eta}=\hat v\nd{\mathcal{Q}}{\eta}\implies\mathcal{C}=\hat v\mathcal{Q}+A,
\end{equation}
for some $A\in\mathbb{R}$. At this point, we introduce the far-field conditions which will substitute the boundary conditions in Eq.~\eqref{nd:bc}. These are
\begin{subequations}\label{TW:bc}
\begin{equation}\label{TW:bc+}
\lim_{\eta\to\infty}\mathcal{C}=\lim_{\eta\to\infty}\mathcal{Q}=0\, ,
\end{equation}
which implies that no contaminant is present nor has been adsorbed far ahead of the travelling wave, and
\begin{equation}\label{TW:bc-}
\lim_{\eta\to-\infty}\mathcal{C}=1,\qquad \lim_{\eta\to-\infty}\mathcal{Q}=q_\text{e}^{*}\, ,
\end{equation}
\end{subequations}
which tells that far behind the travelling wave, the system is in equilibrium.

When applying Eq.~\eqref{TW:bc+} in Eq.~\eqref{TW:relation}, we find $A=0$, whereas using Eq.~\eqref{TW:bc-} determines the travelling wave speed $\hat v=1/\hat q_\text{e}^{*}=1+\kappa^{1/n}>1$.

At this point, only $\mathcal{C}$ remains unknown, but it can be determined via the kinetic equation. In the travelling wave formulation, Eq.~\eqref{nd:sips} becomes
\begin{equation}\label{TW:odeSips}
    -\nd{\mathcal{C}}{\eta}=\mathcal{C}^m\left(1-\frac{\mathcal{C}}{\hat v}\right)^n-
    \left(\frac{\hat v-1}{\hat v}\mathcal{C}\right)^n=:f_{mn}(\mathcal{C}).
\end{equation}
The solutions of Eq.~\eqref{TW:odeSips} have beed discussed in previous works for different combinations of $m,n$ \cite{Aguareles2022,SensiTW} and we will not derive them here again. 

The PFO model reads
\begin{equation}\label{TW:odePFO}
    -\frac{1}{\gamma}\nd{\mathcal{C}}{\eta}=\frac{\mathcal{C}^{\alpha}}{\hat v -1 + \mathcal{C}^{\alpha}}-\frac{\mathcal{C}}{\hat v}=\frac{\hat v\mathcal{C}^{\alpha}-(\mathcal{C}^{\alpha}+\hat v-1)\mathcal{C}}{\hat v(\mathcal{C}^{\alpha}+\hat v-1)}=:g_{\alpha}(\mathcal{C}).
\end{equation}
In contrast to the Sips formulation, the qualitative behaviour of this equation depends on the ratio of $m$ and $n$ rather than on their individual values. The latter will only have an impact on the value of $\hat v=1+\kappa^{1/n}$.

\subsection{Existence of travelling wave solutions}
A travelling wave solution to the problem defined by Eqs. \eqref{nd:masstransfer}, \eqref{nd:pfo} will exist if the solution of Eq.~\eqref{TW:odePFO}, subject to $\mathcal{C}(0)=1/2$, satisfies
\begin{equation}\label{app:proof:conditions}
    \lim_{\eta\to-\infty}\mathcal{C}=1,\qquad \lim_{\eta\to+\infty}\mathcal{C}=0.
\end{equation}
In terms of stability, this condition is equivalent to demanding $\mathcal{C}\equiv1$ to be an unstable equilibrium point of Eq.~\eqref{TW:odePFO}, $\mathcal{C}\equiv0$ must be a stable one. In particular, this requires that $g_\alpha(\mathcal{C})>0$ for $\mathcal{C}\in(0,1)$.

\subsubsection{Case $\alpha\leq1$}\label{app:proof alpha<=1}
Since $g_\alpha(0)=0$ for any $\alpha>0$, we can write
\begin{equation}
    g_{\alpha}(\mathcal{C}) = \frac{\mathcal{C}h_\alpha(\mathcal{C})}{\hat v(\mathcal{C}^{\alpha}+\hat v-1)},
\end{equation}
with
\begin{equation}
    h_\alpha(\mathcal{C}) = \hat{v}\mathcal{C}^{\alpha-1} - \mathcal{C}^\alpha - (\hat{v}-1).
\end{equation}
Since $\hat v>1$, we note that de denominator of $g_\alpha$ never vanishes and thus the non-trivial roots of $g_\alpha$ are contained in $h_\alpha$. Hence, the existence of travelling wave solutions depends on the behaviour of $h_\alpha$. 

In the limit case $\alpha=1$ we have $h_1(\mathcal{C})=1-\mathcal{C}$ and the only non-trivial zero of $g_1$ is $\mathcal{C}=1$. In addition, one has $g_1(\mathcal{C})>0$ for $\mathcal{C}\in(0,1)$.

For $\alpha<1$ we can write 
\begin{equation}
    h_\alpha(\mathcal{C}) = \frac{\hat{v}}{\mathcal{C}^{1-\alpha}} - \mathcal{C}^\alpha - (\hat{v}-1).
\end{equation}
It is trivial to observe that $h_\alpha(1)=0$ and $h_\alpha\to+\infty$ as $\mathcal{C}\to0^+$. Furthermore, we find
\begin{equation}\label{TW:diff_h}
    \nd{h_\alpha}{\mathcal{C}}=-\frac{\hat{v}(1-\alpha)}{\mathcal{C}^{2-\alpha}} - \frac{\alpha}{\mathcal{C}^{1-\alpha}}<0\qquad \forall \mathcal{C}\in(0,1).
\end{equation}
Thus $h_\alpha>0$ and therefore $g_\alpha>0$ for $\mathcal{C}\in(0,1)$, while $g_\alpha(0)=g_\alpha(1)=0$.

\subsubsection{Case $\alpha>1$}\label{app:proof alpha>1}
In this case, we note that
\begin{equation}
    \nd{g_\alpha}{\mathcal{C}} = \frac{\alpha\mathcal{C}^{\alpha-1}(\hat{v} - 1)}{(\mathcal{C}^\alpha + \hat{v} - 1)^2} - \frac{1}{\hat{v}}
\end{equation}
and thus, in particular,
\begin{equation}
    \nd{g_\alpha}{\mathcal{C}}\bigg|_{\mathcal{C}=0}=-\frac{1}{\V}<0,
\end{equation}
which implies that the for any solution of Eq.~\eqref{TW:odePFO}, the steady state $\mathcal{C}=0$ is now unstable.

\begin{figure}
    \centering
    \begin{overpic}[width=.45\textwidth]{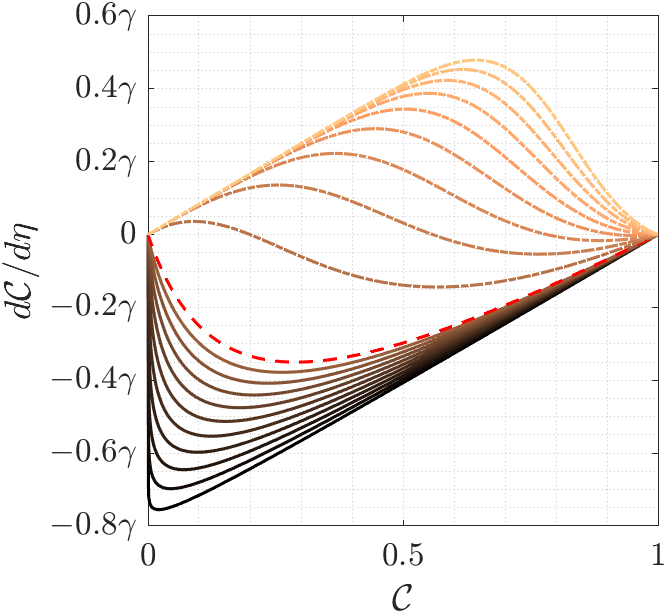}
    \put(91,84){(a)}
    \end{overpic}
    \begin{overpic}[width=.45\textwidth]{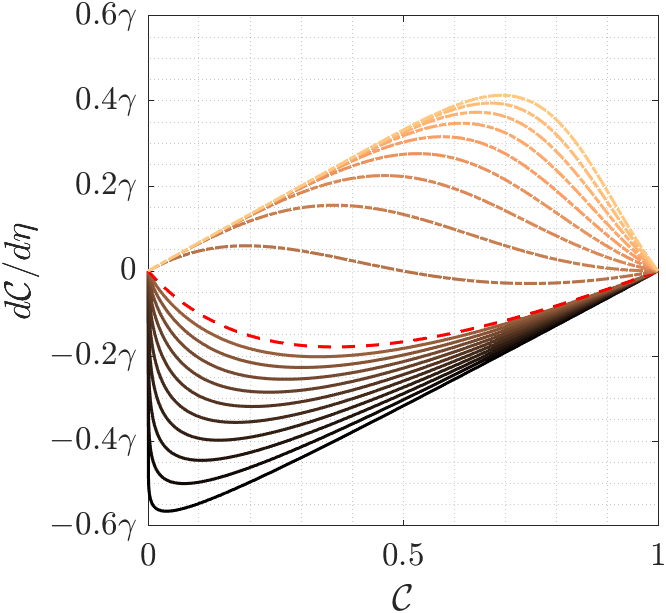}
    \put(91,84){(b)}
    \end{overpic}
    \begin{overpic}[width=.45\textwidth]{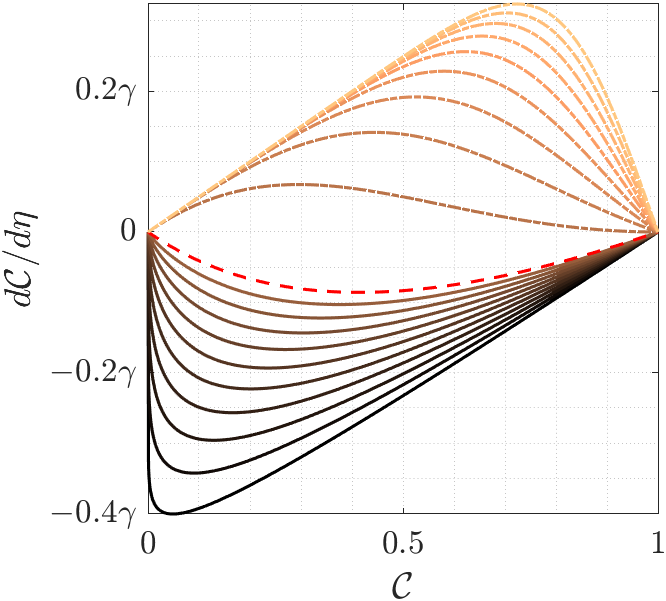}
    \put(91,84){(c)}
    \end{overpic}
    \begin{overpic}[width=.45\textwidth]{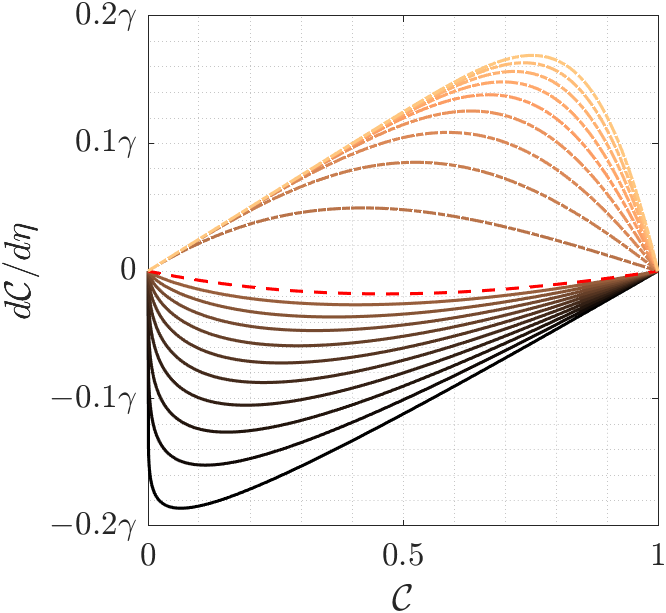}
    \put(91,84){(d)}
    \end{overpic}
    \caption{Phase portraits of Eq.~\eqref{TW:odePFO} for different values of $\alpha\in(0.1,10)$. The case $\alpha=1$ corresponds to the dashed, red line. Solid and dashed dotted lines correspond respectively to $\alpha<1$ and $\alpha>1$. The travelling wave velocities are (a) $\V=1.2$, (b) $\V=1.5$, (c) $\V=2$, (d) $\V=4$.}
    \label{fig:phasep}
\end{figure}

\begin{figure}
    \centering
    \begin{overpic}[width=.49\textwidth]{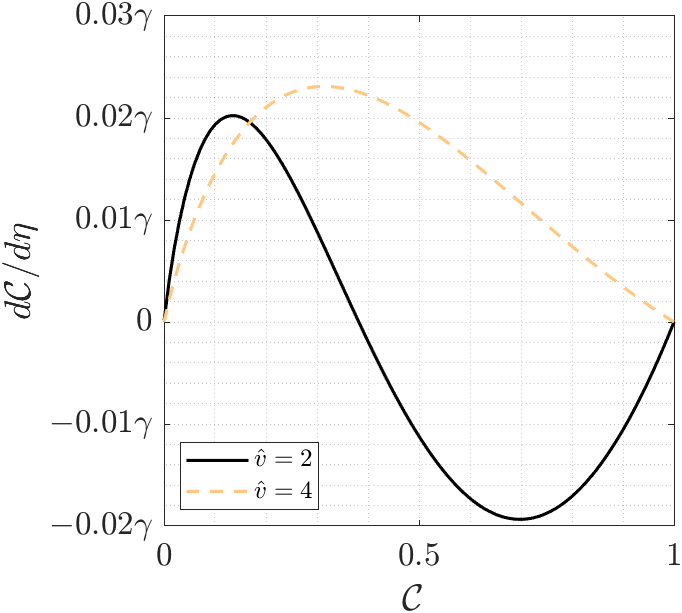}
    \put(90,82){(a)}
    \end{overpic}
    \begin{overpic}[width=.45\textwidth]{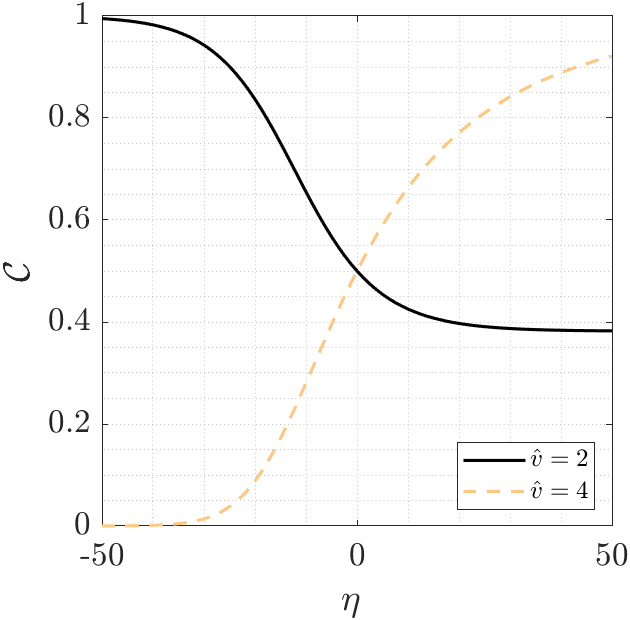}
    \put(18,89){(b)}
    \end{overpic}
    \caption{(a) Phase portrait of Eq.~\eqref{TW:odePFO} for $\alpha=1.5$. (b) Solutions of Eq.~\eqref{TW:odePFO} for $\alpha=1.5$ with $\hat v=2$ and $\hat v=4$.}
    \label{fig:TWalpha>1}
\end{figure}

Figure~\ref{fig:phasep} shows how the behaviour near $\mathcal{C}=0$ changes as soon as $\alpha>1$, for different values of $\V$. Interestingly, we observe that in some cases $g_\alpha$ has an additional root $\mathcal{C}^*\in(0,1)$, implying that the heteroclinic orbit between $\mathcal{C}=1$ and $\mathcal{C}=0$ does not exist anymore. In \ref{app:v alpha} we show that such value $\mathcal{C}^*$ exists only if $\V<\V_\alpha=\alpha/(\alpha-1)$. In Figure~\ref{fig:TWalpha>1} we show the case $\alpha=1.5$ ($\hat v_{1.5}=3$) for two different velocities $\hat v$. It can be clearly observed how in none of the cases the solution satisfies the far-field conditions of Eq.~\eqref{app:proof:conditions}.

\subsection{Implicit solutions for the S-PFO model for $\alpha\leq1$}\label{sec:analyticalslns}
The solution to Eq.~\eqref{TW:odePFO} subject to $\mathcal{C}(0)=1/2$ can be written in implicit form as 
\begin{equation}\label{TW implicit soln general}
    -\gamma\eta =\int_{1/2}^\mathcal{C}\frac{\ud s}{g_{\alpha}(s)}=:\Phi_{\alpha}(\mathcal{C}).
\end{equation}
In line with the analogous analysis done in \cite{Aguareles2022}, we will proceed to determine $\Phi_{\alpha}$ for particular cases of interest.

\subsubsection{Case $\alpha=1$}\label{ssec:alpha=1}
In this case we have $m=n$. In particular, for $m=n=1$ this reduces to the L-PFO model. The function $g_{1}$ is
\begin{equation}
    g_{1}\left(s\right)=\frac{\hat v s-(s+\hat v-1)s}{\hat v(s+\hat v-1)}=\frac{s(1-s)}{\hat v(s+\hat v-1)},
\end{equation}
and hence
\begin{align}
    \begin{split}
        \Phi_{1}(\mathcal{C})=&\hat v\int_{1/2}^\mathcal{C} \frac{s+\hat v-1}{s(1-s)}\ud s\\
        =&\V\int_{1/2}^\mathcal{C}\left(\frac{\V-1}{s}+\frac{\V}{1-s}\right)\ud s\\
        =&\V^2\ln\left(\frac{\mathcal{C}}{1-\mathcal{C}}\right)-\V\ln(2\mathcal{C}).
    \end{split}
\end{align}
Before moving to the next case, we take limits and observe
\begin{equation}\label{TW:Phi1 limits}
    \lim_{\mathcal{C}\to0^+}\Phi_{1}(\mathcal{C})=-\infty\, \qquad \lim_{\mathcal{C}\to1^-}\Phi_{1}(\mathcal{C})=+\infty.
\end{equation}

\subsubsection{Case $\alpha<1$}
In this scenario we observe an important difference with respect to $\alpha=1$. In contrast to the observation made in Eq.~\eqref{TW:Phi1 limits}, here we have
\begin{equation}\label{TW:eta*}
\Phi_\alpha({0^+}):=\lim_{\mathcal{C}\to0^+}\Phi_\alpha(\mathcal{C})>-\infty.
\end{equation}
This property, which we prove in \ref{appB}, is equivalent to the claim that one can find a {finite} cut-off value $\eta^*_\alpha>0$ such that
\begin{equation}
    \eta^*_\alpha=\frac{1}{\gamma}\int_0^{1/2}\frac{\ud s}{g_\alpha(s)}=-\frac{1}{\gamma}\Phi_\alpha(0^+).
\end{equation}
This represents a key qualitative difference with respect to the original Sips formulation.
In the case of the S-PFO model, the individual values of $m$ and $n$ do not matter in the sense of affecting the overall qualitative behaviour of the solutions.

As part of the solution for each of the considered values of $\alpha$, we will also provide the singular value $\eta^*_\alpha$ via Eq.~\eqref{TW:eta*}. Upon recalling the definition of $\eta$, at the outlet ($\X=\hat L$) this leads to the obtention of a breakthrough time $\hat t_b$, defined as
\begin{equation}\label{TW: alpha<1 tb}
    \hat t_b:=\hat t_h-\frac{\eta^*_\alpha}{\V}=\hat t_h-\frac{1}{\gamma\V}\int_0^{1/2}\frac{\ud s}{g_\alpha(s)},
\end{equation}
such that the final breakthrough model defined in Eq.~\eqref{TW:breakthrough} is valid only for $\hat t\geq\hat t_b$, while $\C_b(\T)=0$ for $\T<\T_b$.

\paragraph{Case $\alpha=1/2$} In this case, the function $g_{1/2}$ reads
\begin{equation}
    g_{1/2}\left(s\right)=\frac{\hat vs^{1/2}-(s^{1/2}+\hat v-1)s}{\hat v(s^{1/2}+\hat v-1)} =\frac{s^{1/2}\left(\hat v-(\hat v-1)s^{1/2}-s\right)}{\hat v(s^{1/2}+\hat v-1)}.
\end{equation}
hence we find
\begin{align}
    \begin{split}
        \Phi_{1/2}(\mathcal{C})=&\hat v\int_{1/2}^\mathcal{C} \frac{s^{1/2}+\hat v-1}{s^{1/2}\left(\hat v-(\hat v-1)s^{1/2}-s\right)} \ud s
    \end{split}
\end{align}
Using the change of variable $u=s^{1/2}$ ($ds=2udu$) we find
\begin{subequations}\label{TW:Phi 1/2}
\begin{align}
    \begin{split}
        \Phi_{1/2}(\mathcal{C})=&2\hat v\int_{\sqrt{1/2}}^{\sqrt{\mathcal{C}}} \frac{u+\hat v-1}{\hat v-(\hat v-1)u-u^2} \ud u\\
        =&\frac{2\V}{1+\V}\int_{\sqrt{1/2}}^{\sqrt{\mathcal{C}}} \frac{\V}{1-u}-\frac{1}{\V+u} \ud u\\
        =&\phi_{1/2}({\sqrt{1/2}}) - \phi_{1/2}(\sqrt{\mathcal{C}}),
    \end{split}
\end{align}
where
\begin{equation}
    \phi_{1/2}(u)=\frac{2\V}{1+\V}\bigg[\V\ln(1-u)+\ln(\V+u)\bigg].
\end{equation}
\end{subequations}
In particular, when taking the limit as $\mathcal{C}\to0^+$, we find the cut-off value 
\begin{equation}\label{TW:eta 1/2}
    \eta^*_{1/2} = \frac{2\V}{\gamma(1+\V)}\bigg[\V\ln\left(2+\sqrt{2}\right) - \ln\left(1+\frac{1}{\sqrt{2}\V}\right)\bigg].
\end{equation}

\paragraph{Case $\alpha=1/3$} The function $g_{1/3}$ reads
\begin{equation}
    g_{1/3}\left(s\right)=\frac{\V s^{1/3}-(s^{1/3}+\V -1)s}{\V (s^{1/3}+\V -1)}=\frac{s^{1/3}\left(\V -(\V -1)s^{2/3}-s\right)}{\V (s^{1/2}+\V -1)},
\end{equation}
therefore
\begin{align}
    \begin{split}
        \Phi_{1/3}(\mathcal{C})=&\V \int_{1/2}^\mathcal{C} \frac{s^{1/3}+\V -1}{s^{1/3}\left(\V -(\V -1)s^{2/3}-s\right)} \ud s.
    \end{split}
\end{align}
Similar to the previous case, we introduce the change of variable $u=s^{1/3}$ ($3u^2du=ds$) and find
\begin{align}
    \begin{split}
        \Phi_{1/3}(\mathcal{C})=&3\V \int_{\sqrt[3]{1/2}}^{\sqrt[3]{\mathcal{C}}} \frac{u(u+\V -1)}{\V -(\V -1)u^2-u^3} \ud u\\
        =&3\V \int_{\sqrt[3]{1/2}}^{\sqrt[3]{\mathcal{C}}} \frac{u(u+\V -1)}{(1-u)(\V +\V u +u^2)} \ud u\\
        =&\frac{3\V}{1+2\V}\int_{\sqrt[3]{1/2}}^{\sqrt[3]{\mathcal{C}}} \frac{\V}{1-u}-\frac{(1+\V)u+\V^2}{\V+\V u+u^2} \ud u.
    \end{split}
\end{align}
The first integrand is trivially integrable, but the second depends on the number of real roots of the polynomial $P(u)=\V+\V u+u^2$. We observe that the discriminant is $\Delta=\V(\V-4)$, hence the $P$ has two real roots only if $\V>4$ or, equivalently, $\kappa^{1/n}>3$. In particular, this requires $\kappa>\kappa^{1/n}>3$. From the point of view of the adsorbent, this implies that the adsorbent is expected to perform very poorly, hence the interesting case is the opposite, i.e., when $1<\V<4$ (or $\kappa^{1/n}<3$). In this case, we obtain
\begin{subequations}\label{TW:Phi 1/3}
\begin{align}
        \Phi_{1/3}(\mathcal{C})=&\phi_{1/3}\left(\sqrt[3]{1/2}\right)-\phi_{1/3}\left(\sqrt[3]{\mathcal{C}}\right),
\end{align}
with
\begin{align}
    \begin{split}
        \phi_{1/3}(u)=&\frac{3\V}{1+2\V}\bigg[\V\ln(1-u)+\frac{1+\V}{2} \ln(\V + \hat{v}u + u^2)\\
        &+\frac{\V(\V-1)}{\sqrt{\V(4-\V)}}\arctan\left(\frac{\V+2u}{\sqrt{\V(4-\V)}}\right)\bigg].
    \end{split}
\end{align}
\end{subequations}
Again we can find the cut-off value $\eta^*_{1/3}$ by evaluating $\Phi_{1/3}(0^+)$,
\begin{align}
    \begin{split}\label{TW:eta 1/3}
        \eta^*_{1/3} &= \frac{3\V}{\gamma(1+2\V)} \bigg[ \V\ln\left(\frac{\sqrt[3]{2}}{\sqrt[3]{2}-1}\right) + \frac{1+\V}{2}\ln\left(\frac{\hat{v}\sqrt[3]{4}}{1 + \hat{v}\sqrt[3]{2} + \hat{v}\sqrt[3]{4}}\right) \\
        &- \frac{\V(\V-1)}{\sqrt{\V(4-\V)}} \arctan\left( \frac{\sqrt{4-\V}}{\sqrt{\V}(2\sqrt[3]{2}+1)} \right) \bigg].
    \end{split}
\end{align}

\paragraph{Case $\alpha=2/3$}
The function $g_{2/3}$ takes the form
\begin{equation}
    g_{2/3}\left(s\right)=\frac{\V s^{2/3}-(s^{2/3}+\V -1s}{\V (s^{2/3}+\V -1)}=\frac{s^{2/3}\left(\V -(\V -1)s^{1/3}-s\right)}{\V (s^{2/3}+\V -1)},
\end{equation}
thus 
\begin{align}
    \begin{split}
        \Phi_{2/3}(\mathcal{C})=&\V \int_{1/2}^\mathcal{C} \frac{s^{2/3}+\V -1}{s^{2/3}\left(\V -(\V -1)s^{1/3}-s\right)} \ud s.
    \end{split}
\end{align}
In this case we use again the change of variables $u=s^{1/3}$ ($3u^2du=ds$) to find
\begin{align}
    \begin{split}
        \Phi_{2/3}(\mathcal{C})=&3\V \int_{\sqrt[3]{1/2}}^{\sqrt[3]{\mathcal{C}}} \frac{u^2+\V -1}{\V -(\V -1)u-u^3} \ud u\\
        =&3\V \int_{\sqrt[3]{1/2}}^{\sqrt[3]{\mathcal{C}}} \frac{u^2+\V -1}{(1-u)(\V+u+u^2)} \ud u\\
        =&\frac{3\V}{2+\V}\int_{\sqrt[3]{1/2}}^{\sqrt[3]{\mathcal{C}}} \frac{\V}{1-u}-\frac{2-\V+2u}{\V+u+u^2} \ud u.
    \end{split}
\end{align}
In this case, the polynomial $P(u)=\V+u+u^2$ has no real roots, regardless of the value of $\V>1$. Hence, we find
\begin{subequations}\label{TW:Phi 2/3}
    \begin{equation}
        \Phi_{2/3}(\mathcal{C})=\phi_{2/3}(\sqrt[3]{1/2})-\phi_{2/3}(\sqrt[3]{\mathcal{C}}),
    \end{equation}
    with
    \begin{align}
        \begin{split}
            \phi_{2/3}(u)=&\frac{3\hat{v}}{2+\hat{v}} \bigg[ \hat{v} \ln(1-u) + \ln(u^2 + u + \hat{v}) \\
            &- \frac{2(\hat{v}-1)}{\sqrt{4\hat{v} - 1}} \arctan\left( \frac{2u + 1}{\sqrt{4\hat{v} - 1}} \right) \bigg],
        \end{split}
    \end{align}
\end{subequations}
and hence the cut-off $\eta^*_{2/3}$ is
\begin{align}
    \begin{split}\label{TW:eta 2/3}
        \eta^*_{2/3}=&\frac{3\hat{v}}{\gamma(\hat{v} + 2)}\bigg[ \V \ln\left( \frac{\sqrt[3]{2}}{\sqrt[3]{2}-1} \right) + \ln\left( \frac{\V}{\V\sqrt[3]{4} + \sqrt[3]{2} + 1} \right) \\
        &+ \frac{2(\hat{v} - 1)}{\sqrt{4\hat{v} - 1}} \arctan \left( \frac{\sqrt{4\hat{v} - 1}}{2\hat{v}\sqrt[3]{2} + 1} \right) \bigg].
    \end{split}
\end{align}

\subsection{TW solution of the LDF model}
When neglecting desorption (i.e., setting $\kappa=0$) we obtain $\V=1$ and Eq.~\eqref{TW:odePFO} reduces to
\begin{equation}\label{TW:LDF no desorption}
    \nd{\mathcal C}{\eta}=\gamma(\mathcal{C}-1)\implies\mathcal{C}(\eta)= 1-\frac{e^{\gamma\eta}}{2},
\end{equation}
where we have used the initial condition $\mathcal{C}(0)=1/2$. We note that again we have a cut-off value $\eta_0^*$, given by
\begin{equation}
    \eta_0^*=\frac{\ln(2)}{\gamma}.
\end{equation}

\section{Results}\label{sec:5}

\subsection{Cut-off value $\eta^*_\alpha$}
\begin{figure}
    \centering
    \centering
    \begin{overpic}[width=.48\textwidth]{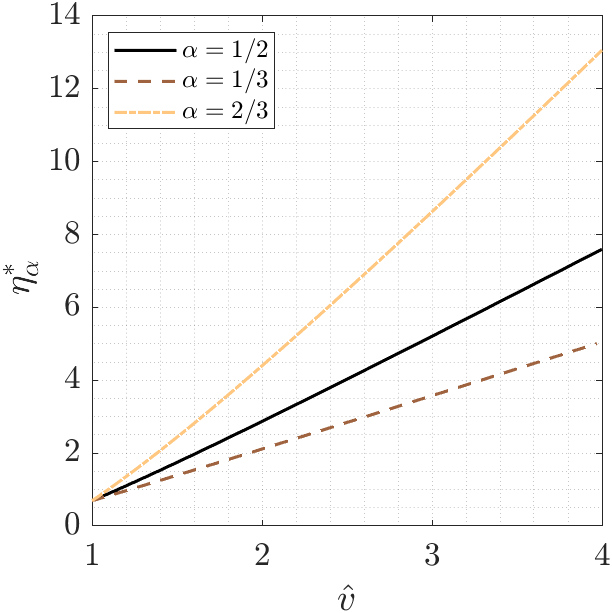}
    \put(90,92){(a)}
    \end{overpic}
    \begin{overpic}[width=.49\textwidth]{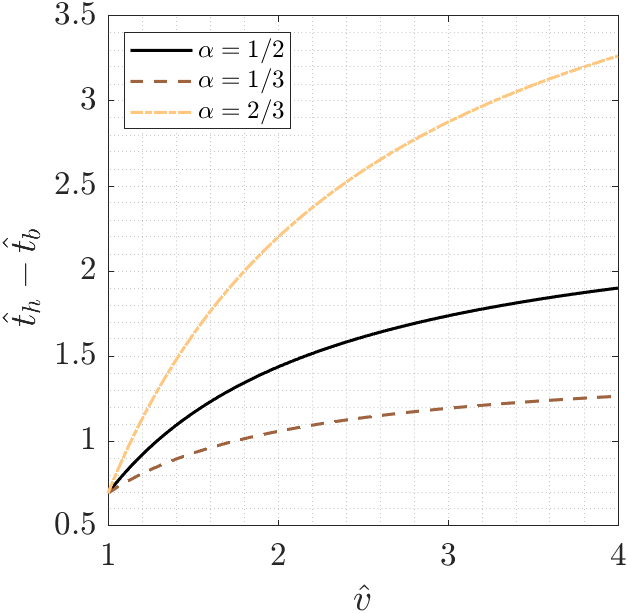}
    \put(90,90){(b)}
    \end{overpic}
    \caption{Dependence of (a) $\eta^*_\alpha$ and (b) $\hat t_h-\hat t_b$ as functions of the travelling wave velocity $\V$. In both panels we have fixed $\gamma=1$.}
    \label{fig:eta alpha}
\end{figure}

\begin{figure}
    \centering
    \begin{overpic}[width=.5\textwidth]{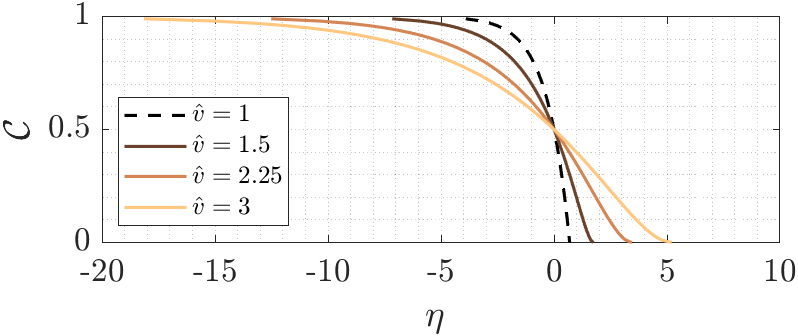}
    \put(90,35){(a)}
    \end{overpic}
    \begin{overpic}[width=.5\textwidth]{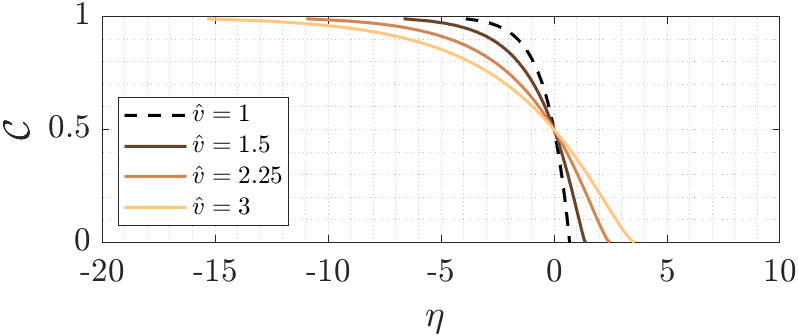}
    \put(90,35){(b)}
    \end{overpic}
    \begin{overpic}[width=.5\textwidth]{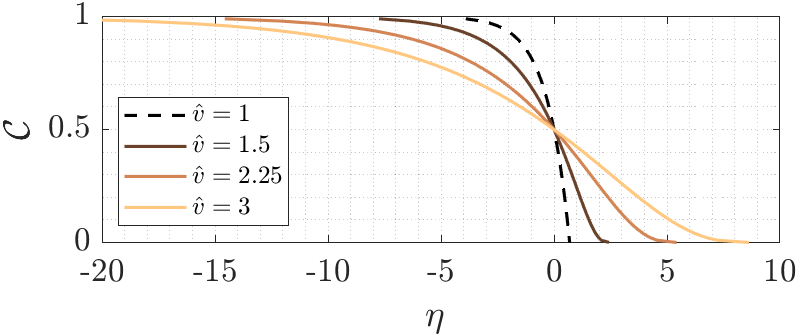}
    \put(90,35){(c)}
    \end{overpic}
    \caption{Travelling wave solutions for (a) $\alpha=1/2$, (b) $\alpha=1/3$, (c) $\alpha=2/3$. The limiting case with no desorption ($\hat v=1$), where all the models collapse to the same profile (Eq.~\eqref{TW:LDF no desorption}) is shown for comparison.}
    \label{fig:TW vs v}
\end{figure}

Contrary to the case $\alpha=1$ or the Sips model for $m\leq n$ with $m,n\geq1$, the breakthrough curve for $\alpha<1$ presents a sudden breakthrough at a breakthrough time $\T_b$. As it can be observed in Eq.~\eqref{TW: alpha<1 tb}, the cut-off value $\eta^*_\alpha$ represents essentially the time gap between the breakthrough time and the half-time, i.e., it is a measure of how fast the adsorbate concentration is increasing at the outlet. Since $\eta_\alpha^*\sim1/\gamma$, for large values of $\gamma$ we expect a very fast increase, whereas the breakthrough slows down as $\gamma$ decreases. In Figure~\ref{fig:eta alpha}a we show how $\eta_{1/2}^*,\eta_{1/3}^*,\eta_{2/3}^*$ vary as functions of $\V$. We observe that the values of $\eta_\alpha$ increase as $\V$ increases, which at first sight does not align with the fact that it represents the gap between breakthrough and half-times. However, recall that $\eta_\alpha^*=-\V(\hat t_b-\hat t_h)$, hence the real gap is given by $\eta_\alpha^*/\V$. The latter is shown in Figure~\ref{fig:eta alpha}b, where we can observe how the gap still increases as $\V$ increases, but in a concave rather than convex fashion. The effet on the travelling wave profiles is shown in Figure~\ref{fig:TW vs v}. The increase of the gap can be explained by recalling $\V=1+\kappa^{1/n}=q_\text{max}/q_\text{e}^{*}$: if an adsorbent is not capable of removing large amounts of adsorbate ($q_\text{e}^*\ll q_\text{max}$), then one expects the breakthrough to happen earlier. Alternatively, if the adsorbent also desorbs at a high rate (increase $\kappa$), then the contaminant reaches the outlet  faster as opposed to having less desorption.

\subsection{Comparison of travelling wave solutions}
\begin{figure}
    \centering
    \begin{overpic}[width=.3\textwidth]{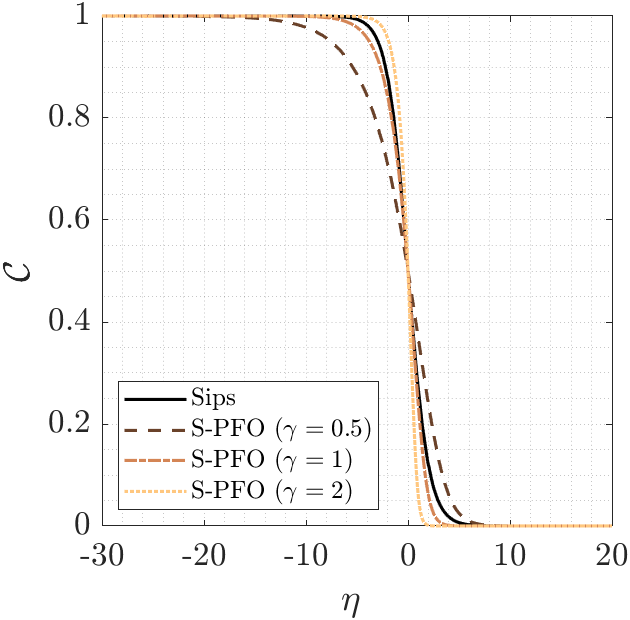}
    \put(80,85){(1,1)}
    \end{overpic}
    \begin{overpic}[width=.3\textwidth]{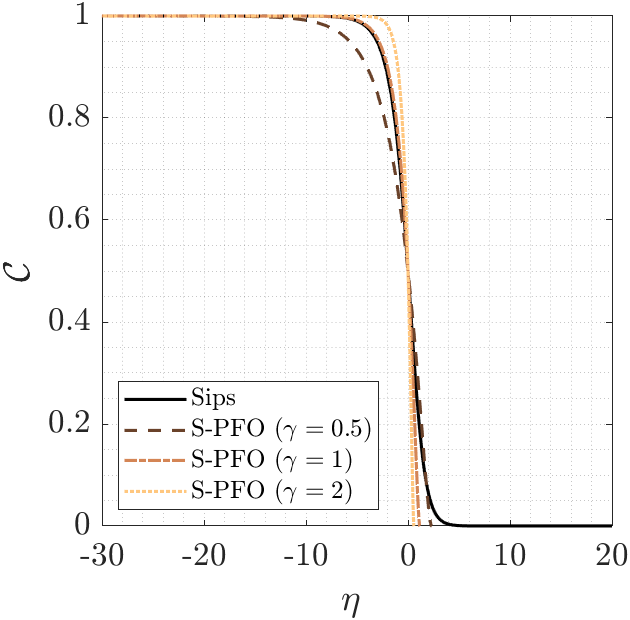}
    \put(80,85){(1,2)}
    \end{overpic}
    \begin{overpic}[width=.3\textwidth]{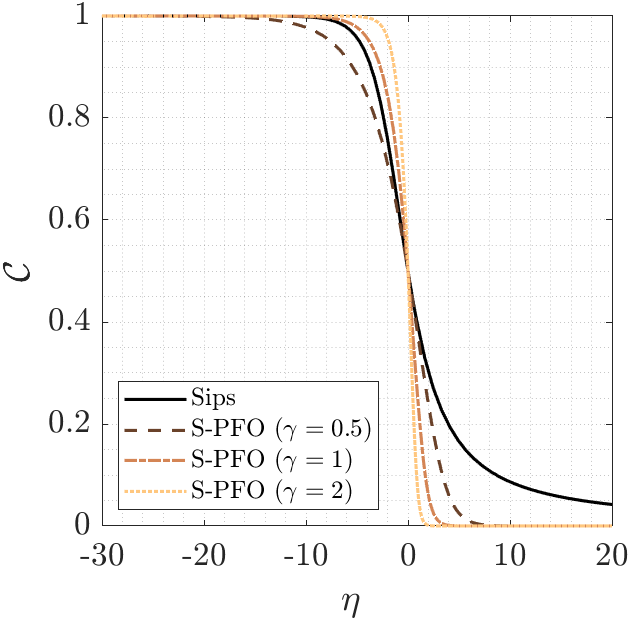}
    \put(80,85){(2,2)}
    \end{overpic}
    \begin{overpic}[width=.3\textwidth]{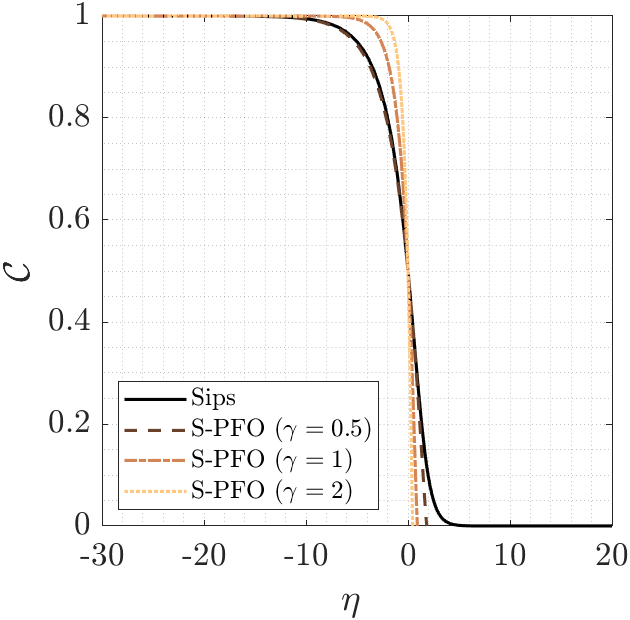}
    \put(80,85){(1,3)}
    \end{overpic}
    \begin{overpic}[width=.3\textwidth]{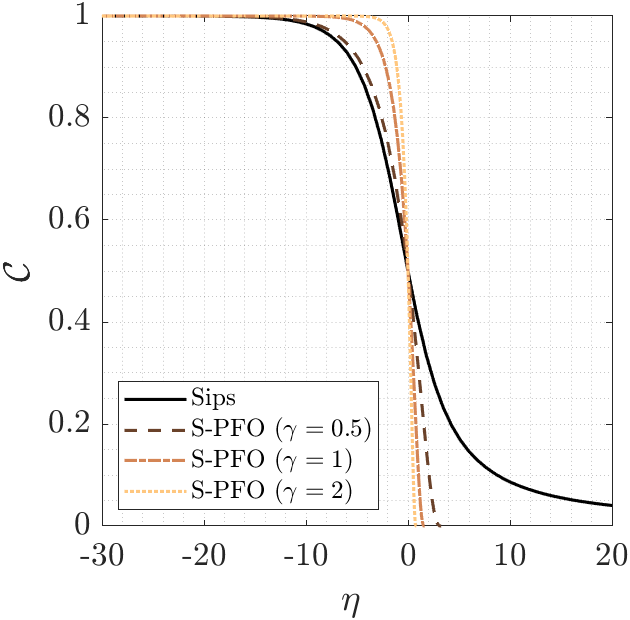}
    \put(80,85){(2,3)}
    \end{overpic}
    \begin{overpic}[width=.3\textwidth]{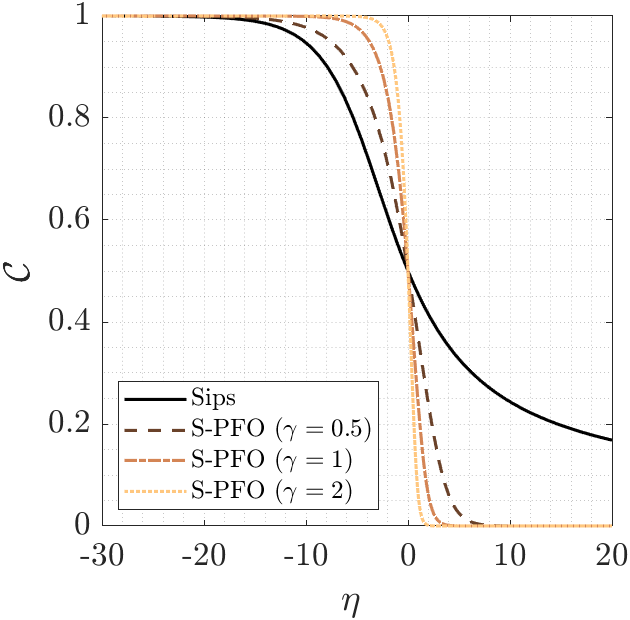}
    \put(80,85){(3,3)}
    \end{overpic}
    \caption{Comparison of the solutions to Eqs.~\eqref{TW:odeSips} and \eqref{TW:odePFO}, with $\hat v=3/2$ (equivalently $\Q_\text{e}=2/3$). The value of $m$ and $n$ is specified in the top right corner of each panel.
    }
    \label{fig:TW_Comp}
\end{figure}

In this section we compare the solutions of Eqs.~\eqref{TW:odeSips} and \eqref{TW:odePFO}. Furthermore, the latter is solved for 3 different values of $\gamma$ to assess its impact on the solution. The results are shown in Fig~\ref{fig:TW_Comp}. Let us discuss to cases separately. 

The parameter $\gamma$ works as a scaling factor. For $\gamma\ll1$ the breakthrough curves of S-PFO and Sips models coincide towards the end of the process, whereas for $\gamma\gg1$ the curves are similar after the initial breakthrough, but diverge as $\hat t\gg1$. Except for the case when $m=n$ (i.e., $\alpha=1$), breakthrough for the S-PFO model occurs abruptly at $\hat t=\hat t_b$, in contrast to the smooth increase from the Sips model. As $\V\to1$, we note that all the cut-off values tend to the same value, which is precisely the cut-off value $\eta_0^*=\ln(2)$ of the LDF model with no desorption.

\subsection{Dimensional breakthrough models}
In terms of the original non-dimensional variables and setting $\hat x=\hat L$, Eq.~\eqref{TW implicit soln general} becomes
\begin{equation}\label{TW:breakthrough}
    \hat t-\hat t_h =\frac{1}{\gamma\V}\Phi_{\alpha}(\hat c_b),
\end{equation}
where $\hat c_b=\hat c(\hat L,\hat t)$.  Reverting the non-dimensional formulation in Eq.~\eqref{TW:breakthrough}, we obtain the final breakthrough model in implicit form, 
\begin{equation}\label{TW:breakthroughDIM}
    t-t_h =\frac{1}{k_{LDF}(1+\kappa^{1/n})}\Phi_{\alpha}(c_b/c_\text{in}),
\end{equation}
with $\kappa=k_d/(k_ac_\text{in}^m)$. For $m<n$ (equivalently $\alpha<1$), this implicit form is only valid for $t\geq t_b$, where the breakthrough time is 
\begin{equation}
    t_b=t_h-\frac{\Phi_\alpha(0^+)}{k_{LDF}(1+\kappa^{1/n})}.
\end{equation}
Conversely, for $t<t_b$ we have $c_b(t)=0$. In contrast, the breakthrough models derived for the Sips model in \cite{Aguareles2022} were of the form
\begin{equation}\label{TW:breakthroughDIMSips}
    t-t_h =\frac{1}{k_ac_\text{in}^mq_\text{max}^{n-1}(1+\kappa^{1/n})}\Psi_{mn}(c_b/c_\text{in}),
\end{equation}
valid for any $t>0$ (provided $m,n\geq1$).

\subsection{Limiting cases}

\begin{figure}
    \centering
    \begin{overpic}[width=.45\textwidth]{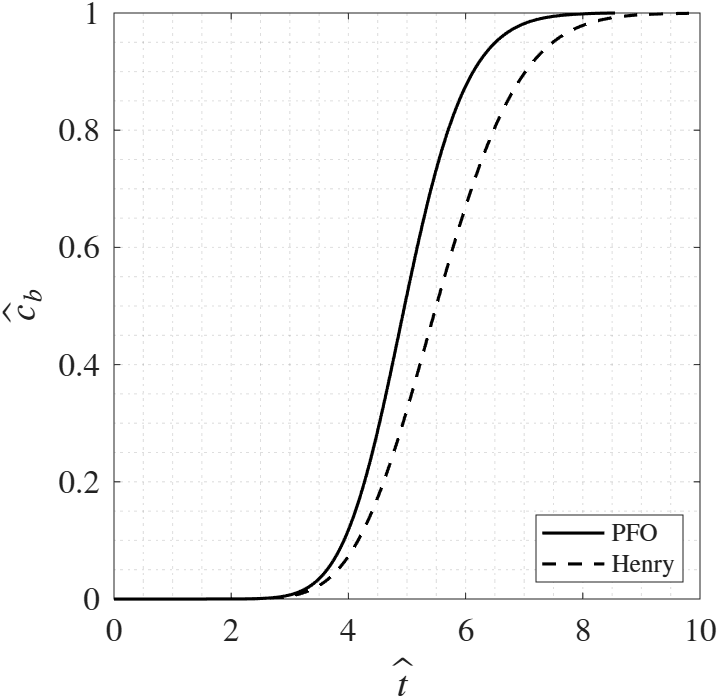}
    \put(89,89){(a)}
    \end{overpic}
    \begin{overpic}[width=.44\textwidth]{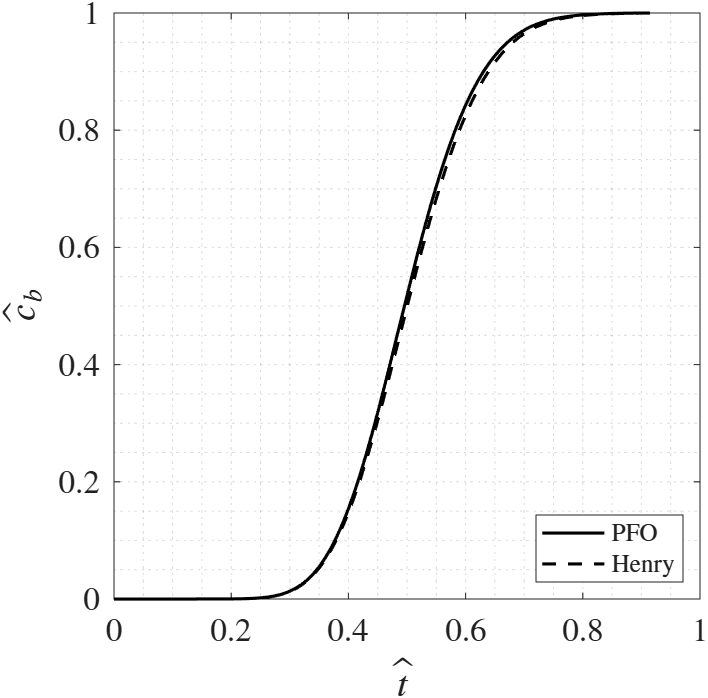}
    \put(89,89){(b)}
    \end{overpic}
    \caption{Breakthrough curves for the limit $K_S c_\text{in}\ll1$, computed numerically. Dashed lines correspond to the S-PFO model with $m=n=1$, whereas solid lines correspond to the Henry model, Eq.~\eqref{eqHenry}. In both panels, we have set $\Da=10^{-5}$, $\Pe^{-1}=0.01$, $\gamma=\kappa$ and (a) $\kappa=9$ and (b) $\kappa=99$.}
    \label{fig:Henry}
\end{figure}

Throughout this work, we have developed the solutions for the S-PFO model and discussed the differences with respect to those of the original Langmuir/Sips formulation. However, in certain limits, both formulations can lead to similar solutions. Here we show that for $n,m=1$ there are two limits in which both Eqs.~\eqref{eq:sips} and \eqref{eq:LDFdynamic} reduce to similar expressions. For inefficient adsorbents (large desoprtion) or extremely low concentrations of contaminants, we can assume $k_d\gg k_a c_\text{in}$, i.e., $K_Sc_\text{in}\ll1$. Under this assumption, the isotherm provided in Eq.~\eqref{eq:isosips} reduces to
\begin{equation}
    q_\text{e}\approx K_Sc_\text{in}q_\text{max}\ll q_\text{max}.
\end{equation}
In particular, one also expects $q\ll q_\text{max}$, which leads to Eq.~\eqref{eq:sips} being reduced to 
\begin{equation}\label{eqHenry}
    F(c,q)\approx k_d\left(K_Sq_\text{max}c-q\right),
\end{equation}
which is the well known Henry model \cite{McCabe1993}.
The L-PFO equation can be reduced to a similar form. Since $K_Sc\leq K_Sc_\text{in}\ll1$, Eq.~\eqref{eq:LDFdynamic} can be reduced to
\begin{equation}
    F(c,q)\approx k_{LDF}\left(K_Sq_\text{max}c-q\right).
\end{equation}
Hence, for $K_Sc_\text{in}\ll1$, both models reduce to the dynamic Henry model, each with its own kinetic rate $k_d$ or $k_{LDF}$. In Figure~\ref{fig:Henry} we show how reducing $K_Sc_\text{in}\to0$ (that is, taking $\kappa\to\infty$), leads to the L-PFO model effectively reproducing the Henry model. Note that we have set $k_d=k_{LDF}$ (hence $\gamma=\kappa$ in the non-dimensional version).

\subsection{Validation against experimental data}

In order to assess the adequacy of S-PFO and Sips models to describe real experimental results, we consider isotherm and breakthrough experimental data of different adsorption applications. The details of the different datasets considered are provided in Table~\ref{Tab:data}.

\begin{table}[H]
\centering
\caption{Experimental operating conditions for the various applications investigated in this study from Myers et al.\cite{Myers23} (dataset 1), Sulaymon et al. \cite{Sulaymon2009} (dataset 2) and Sulaymon et al. \cite{Sulaymon2014} (dataset 3) for Toluene (C$_7$H$_8$), Copper(II) (Cu$^{2+}$) and Mercury(II) (Hg$^{2+}$) respectively. SAC: steam activated carbon; GAC: granular activated carbon; DAS: dry activated sludge.}\label{Tab:data}
\smallskip
\begin{tabular}{@{}lccccc@{}}
\toprule
 & \textbf{Dataset 1}~\cite{Myers23} & \textbf{Dataset 2}~\cite{Sulaymon2009} & \textbf{Dataset 3}~\cite{Sulaymon2014}\\
\midrule
Solute & Toluene & Cu(II) & Hg(II) \\
Carrier & N$_2$ & H$_2$O &H$_2$O  \\
Temperature (K) & 299.15 & 298.15 & 298.15 \\
Pressure (bar) & 1.12 & 1 & 1 \\
Adsorbent & SAC & GAC & DAS\\
Mass (kg) & $1.6 \times 10^{-4}$ & 0.038 & 0.077\\
Bed length (m) & 0.0054 & 0.1 & 0.05\\
Internal diameter (m) & 0.01 & 0.0381 & 0.050\\
Flow rate (L/min) & 0.205 & 0.06 & 0.084\\
Particle diameter (m) & $3.2 \times 10^{-3}$& $6 \times 10^{-4}$ & $3.8 \times 10^{-4}$  \\
Bulk density (kg/m$^3$) & 377.3  & 336 & 784.4 \\
Inlet concentration (mol/m$^3$) & 0.004--0.031  & 0.39--1.18 & 0.13--0.50 \\
\bottomrule
\end{tabular}
\end{table}

The analytical breakthrough expressions derived in Section~\ref{sec:4} were fitted to the experimental data to calibrate the only adjustable parameter in each model: $k_a$ for the Sips model and $k_{LDF}$ for the S-PFO model. Both models incorporate the maximum adsorption capacity, $q_{\text{max}}$. In addition, the S-PFO model includes the equilibrium constant, $K_S$, whereas the Sips model includes the desorption rate coefficient, $k_d$, which is related to $K_S$ through Eq.~\eqref{eq:isosips}. However, both $q_{\text{max}}$ and $K_S$ can be determined independently by fitting the equilibrium experimental data to the Sips isotherm given in Eq.~\eqref{eq:isosips}. Details on the calculation of the experimental equilibrium adsorption capacities, $q_e^{\mathrm{exp}}$, together with the results of the isotherm fitting and the corresponding fitted parameters, are provided in \ref{sec:appiso}.

Once the isotherm parameters have been determined, the remaining kinetic parameters, $k_a$ for the Sips model and $k_{LDF}$ for the S-PFO model, are estimated by fitting the functions $\Psi_{mn}$ and $\Phi_\alpha$, respectively, using the experimental breakthrough concentration data. According to Eq.~\eqref{TW:breakthroughDIMSips} and Eq.~\eqref{TW:breakthroughDIM}, these functions are expected to vary linearly with $t-t_h$. A detailed description of the fitting procedure is provided in \ref{sec:appfittingprocedure}.

\begin{figure}[H]
    \centering
    \begin{overpic}[width=.45\textwidth]{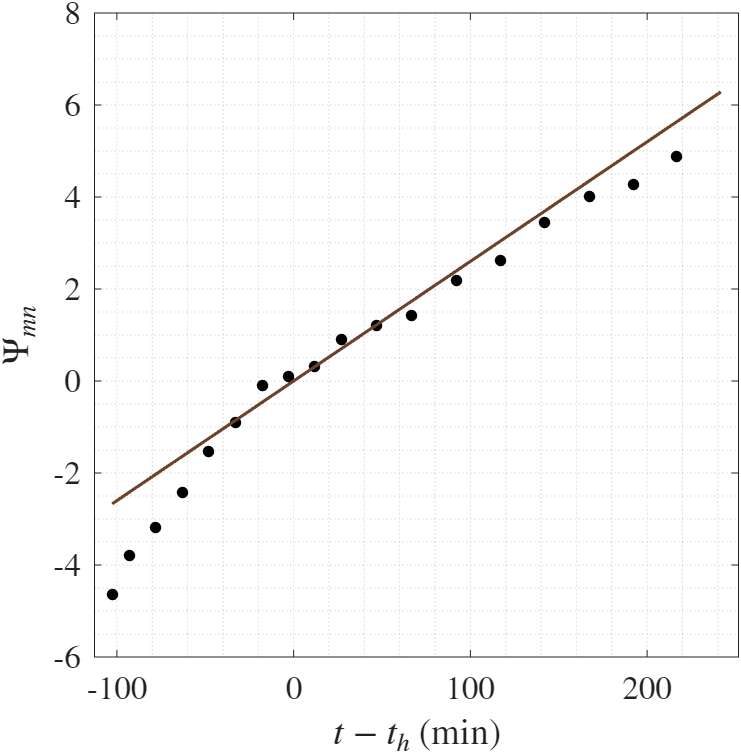}
    \put(86,16){(1,1)}
    \end{overpic}
    \begin{overpic}[width=.45\textwidth]{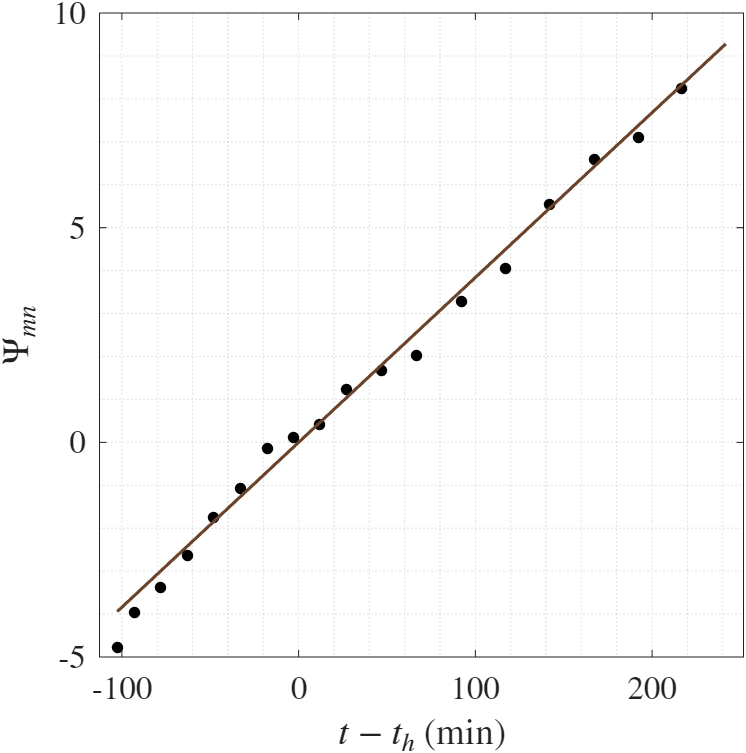}
    \put(86,16){(1,2)}
    \end{overpic}
    \begin{overpic}[width=.45\textwidth]{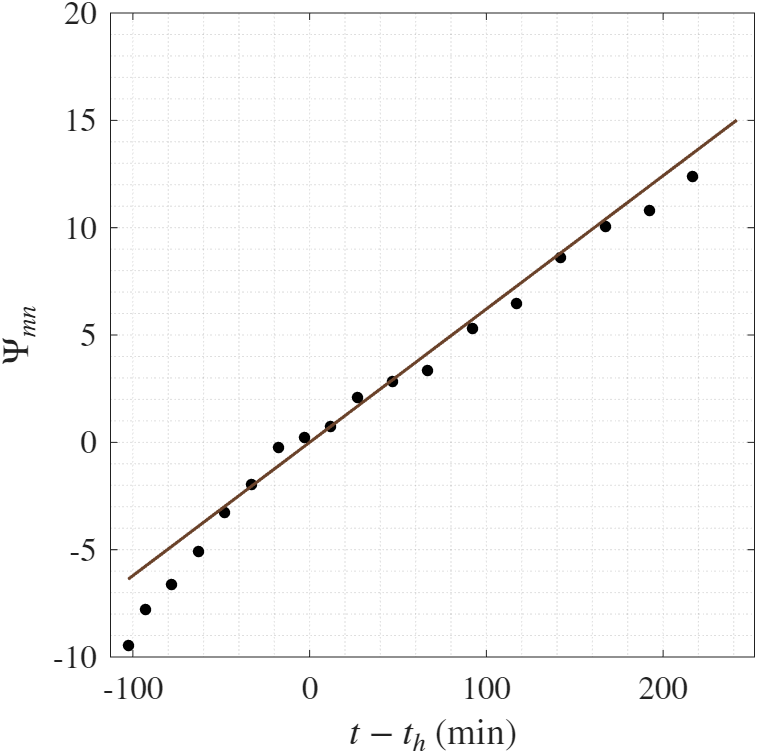}
    \put(86,16){(1,3)}
    \end{overpic}
    \begin{overpic}[width=.45\textwidth]{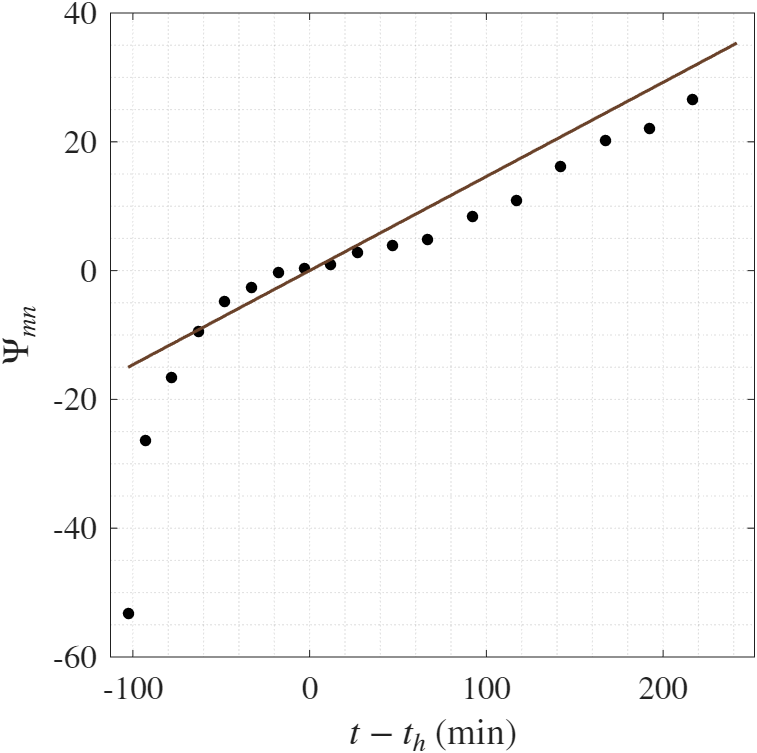}
    \put(86,16){(2,3)}
    \end{overpic}
    \caption{Linear fitting of $\Psi_{mn}$ vs $t-t_h$ for dataset 2 \cite{Sulaymon2009} (operating conditions in Table~\ref{Tab:data}) with inlet concentration $c_{in}=0.3934$ mol/m$^3$ for different $(m,n)$ combinations.}
    \label{fig:psiVSt-t12}
\end{figure}

An example of the linear fitting procedure is presented in Figure~\ref{fig:psiVSt-t12}. Each panel shows the fit of the Sips model \cite{Aguareles2022} to the experimental data reported by Sulaymon et al. \cite{Sulaymon2009} using Eq.~\eqref{eq:fitting} for different combinations of $(m,n)$. Among the cases considered, only $(m,n)=(1,2)$ exhibits a clear linear trend. This is consistent with the results of the isotherm analysis presented in \ref{sec:appiso}, where the Sips isotherm with $\alpha=1/2$ was found to provide an excellent description of the equilibrium data. 

As shown in Figure~\ref{fig:isotherms} and Table~\ref{tab:appgofiso}, the adsorption isotherms do not provide sufficient evidence to identify a unique optimal value of $\alpha$. For each of the three datasets, at least two different configurations yield comparably good fits to the equilibrium data. Therefore, the selection of the appropriate $(m,n)$ values for the experimental datasets listed in Table~\ref{Tab:data} was based primarily on their ability to reproduce the breakthrough curves, while ensuring consistency with the corresponding isotherm analysis. In all cases, the optimal $(m,n)$ configuration for the Sips breakthrough model coincides with the optimal $\alpha$ value for the S-PFO breakthrough model, as illustrated in Figure~\ref{fig:appBCmn}. 

The final isotherm parameters used for the breakthrough analysis are provided in Table~\ref{tab:isotherm_parameters}. The identification of dataset 1 with $(m,n)=(1,1)$ (Langmuir kinetics) is consistent with several studies reporting that the adsorption of toluene onto activated carbon is predominantly governed by physical adsorption \cite{Tran2026,Yang2018,Myers23}.

For divalent metal ions such as Cu(II), adsorption onto activated carbon has been described as a complexation process involving one metal ion and two carboxylic groups on the adsorbent surface \cite{BenAmar2024,Soria2020,Xu2016}. Such a mechanism is consistent with a reaction scheme of $(m,n)=(1,2)$. Similarly, complexation through hydroxyl groups in the polysaccharide and protein components that make up most of the dry activated sludge (DAS) cell wall has been identified as the primary adsorption mechanism for Hg(II) \cite{Yousif2013,Sulaymon2014}. The contribution of additional adsorption mechanisms, such as ion exchange \cite{Yousif2013,Sulaymon2014}, electrostatic attraction \cite{Vanveenhuyzen2021,Giraldo2020}, and internal cell adsorption \cite{Deng2026,He2024}, may result in an overall reaction order of $(m,n)=(1,3)$.

\begin{figure}[H]
    \centering
    \begin{overpic}[width=.45\textwidth]{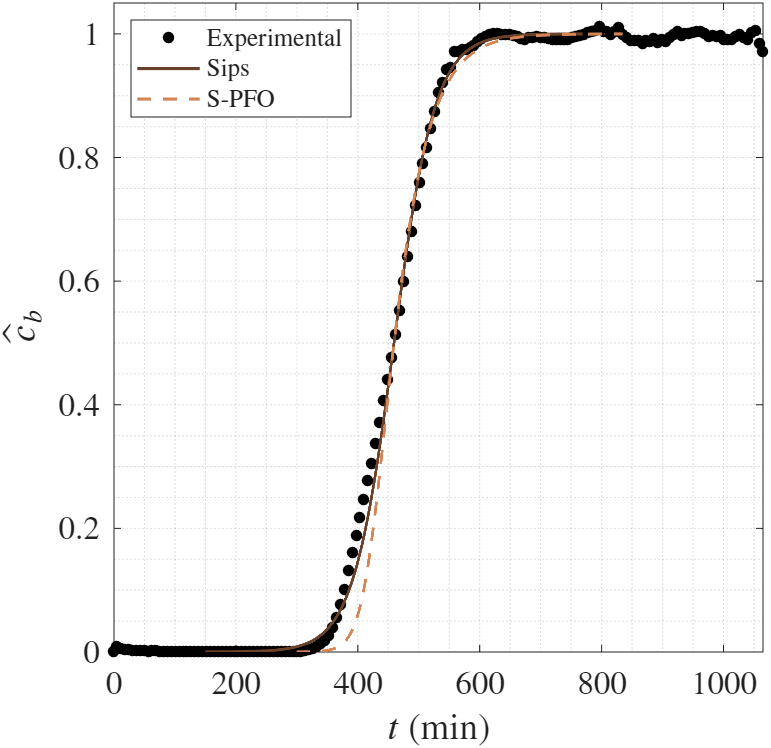}
    \put(90,16){(a)}
    \end{overpic}
    \begin{overpic}[width=.45\textwidth]{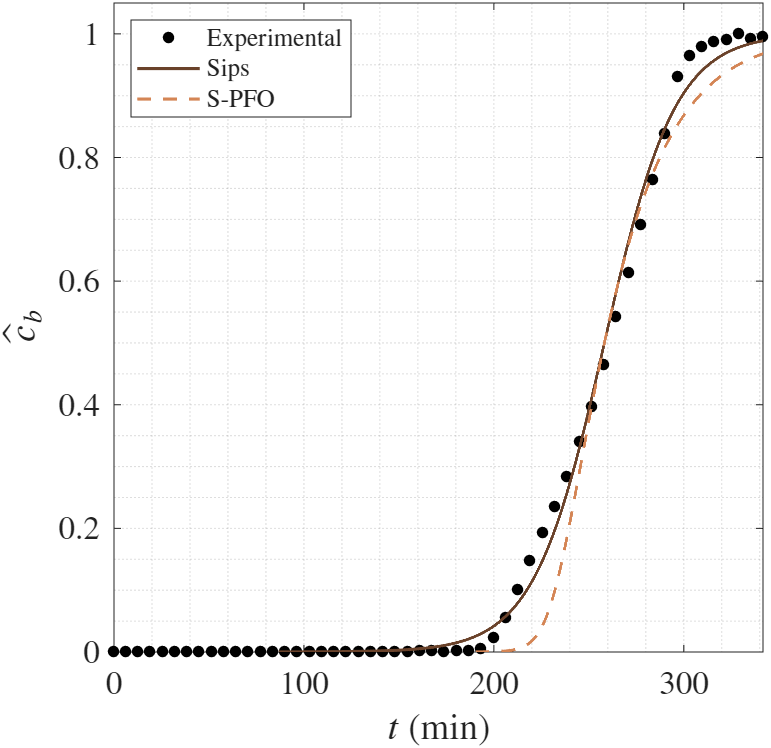}
    \put(90,16){(b)}
    \end{overpic}
    \begin{overpic}[width=.45\textwidth]{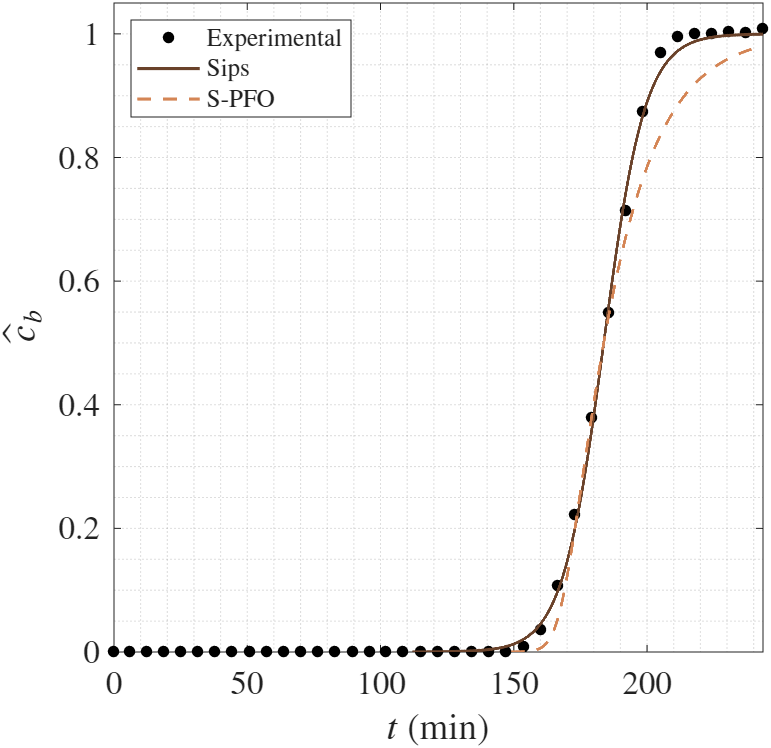}
    \put(90,16){(c)}
    \end{overpic}
    \begin{overpic}[width=.45\textwidth]{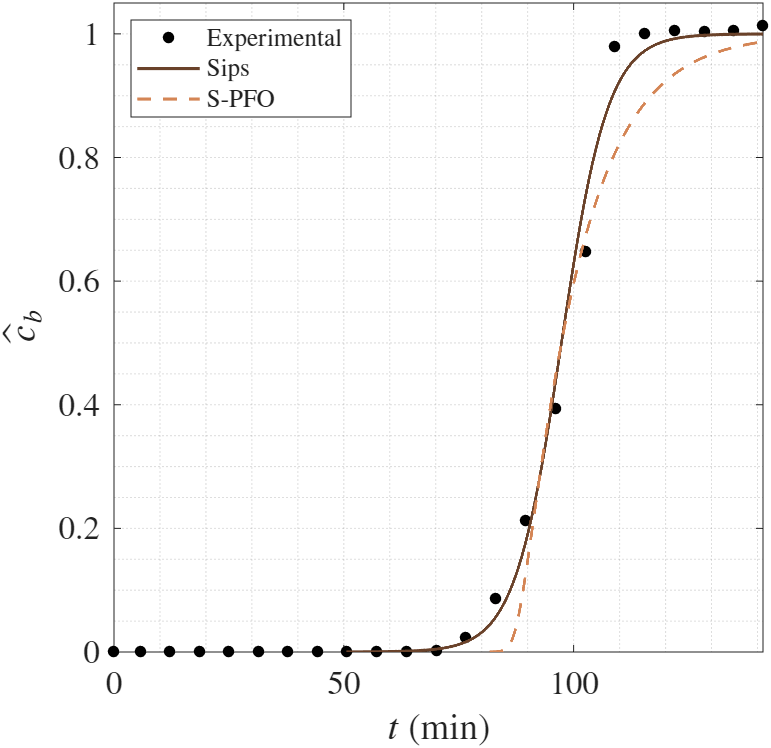}
    \put(90,16){(d)}
    \end{overpic}
    \caption{Breakthrough curves obtained for both Sips and S-PFO models with $(m,n)=(1,1)$ for different inlet concentrations, applied to experimental dataset 1 for the adsorption of toluene onto steam activated carbon \cite{Myers23} (operating conditions in Table~\ref{Tab:data}). (a) $c_{in}=4.44\times10^{-3}$ mol/m$^3$, (b) $c_{in}=1.04\times10^{-2}$ mol/m$^3$, (c) $c_{in}=1.43\times10^{-2}$ mol/m$^3$, (d) $c_{in}=3.08\times10^{-2}$ mol/m$^3$.}
    \label{fig:TolRB3}
\end{figure}

\begin{figure}[H]
\begin{center}
    \begin{overpic}[width=.33\textwidth]{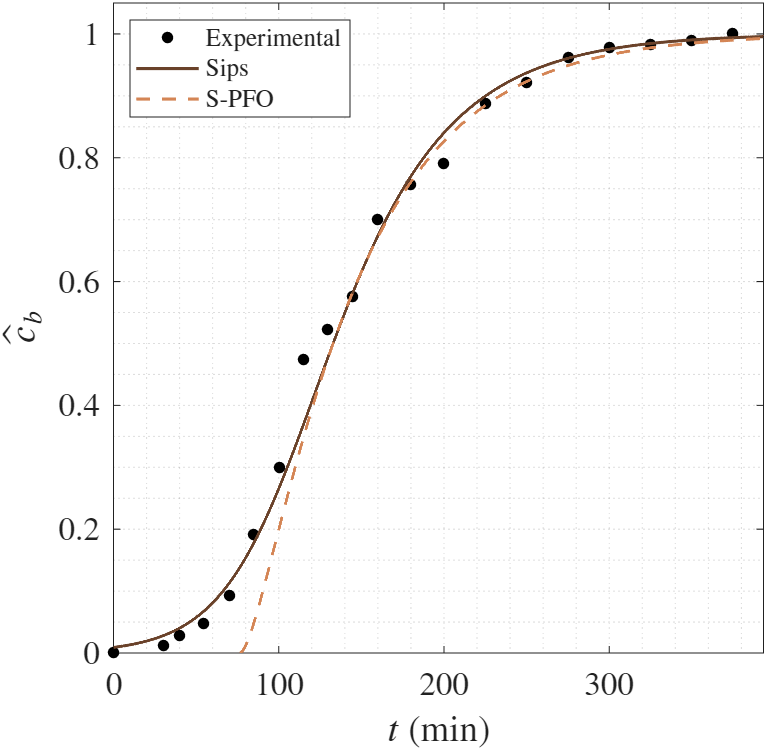}
    \put(89,16){(a)}
    \end{overpic}%
    \begin{overpic}[width=.33\textwidth]{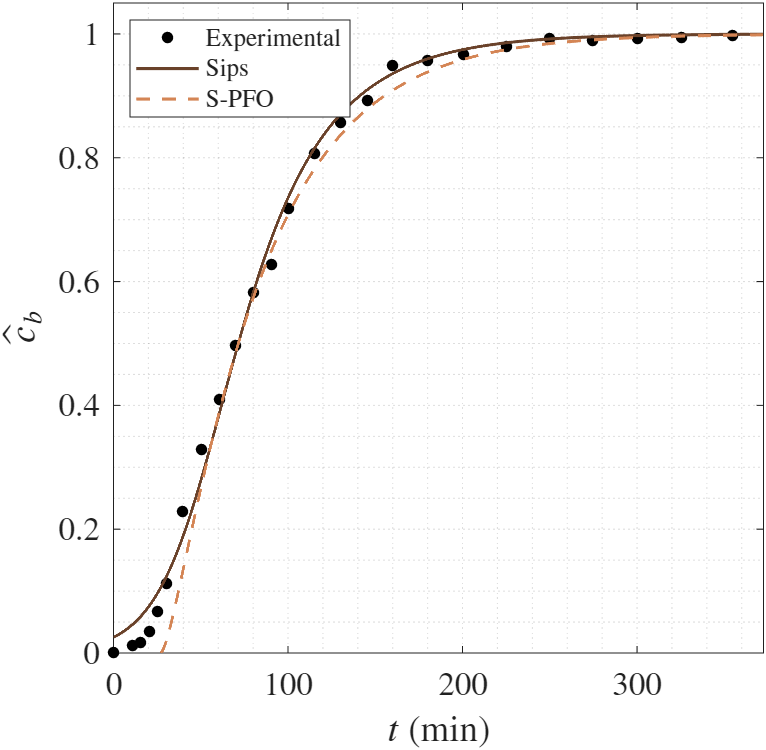}
    \put(89,16){(b)}
    \end{overpic}%
    \begin{overpic}[width=.33\textwidth]{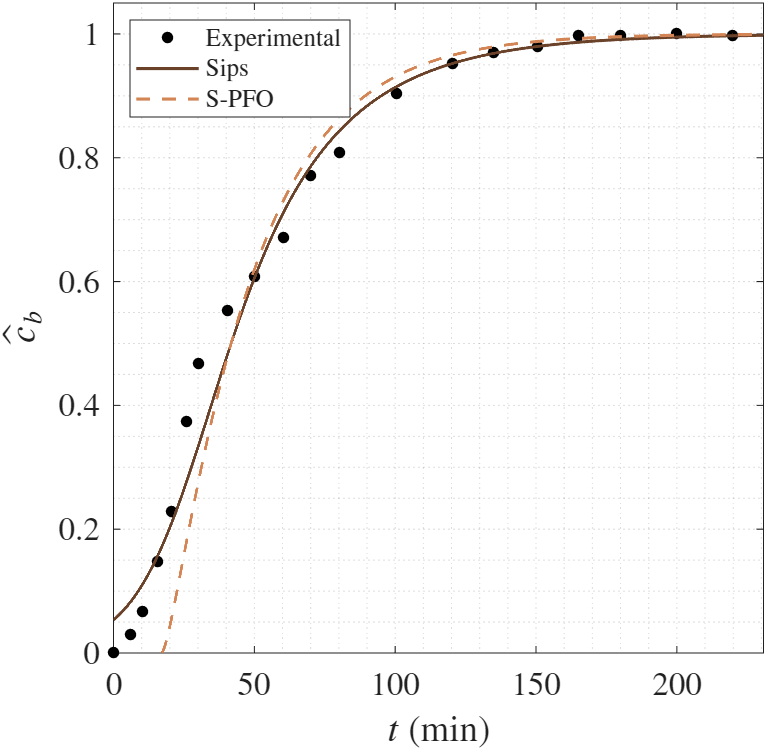}
    \put(89,16){(c)}
    \end{overpic}
\end{center}
    
    \caption{Breakthrough curves obtained for both Sips and S-PFO models with $(m,n)=(1,2)$ for different inlet concentrations, applied to experimental dataset 2 for the adsorption of copper(II) onto granular activated carbon \cite{Sulaymon2009} (operating conditions in Table~\ref{Tab:data}). (a) $c_{in}=0.39$ mol/m$^3$, (b) $c_{in}=0.79$ mol/m$^3$, (c) $c_{in}=1.18$ mol/m$^3$.}
    \label{fig:SulaymonCEJ}
\end{figure}

\begin{figure}[H]
    \centering
    \begin{overpic}[width=.45\textwidth]{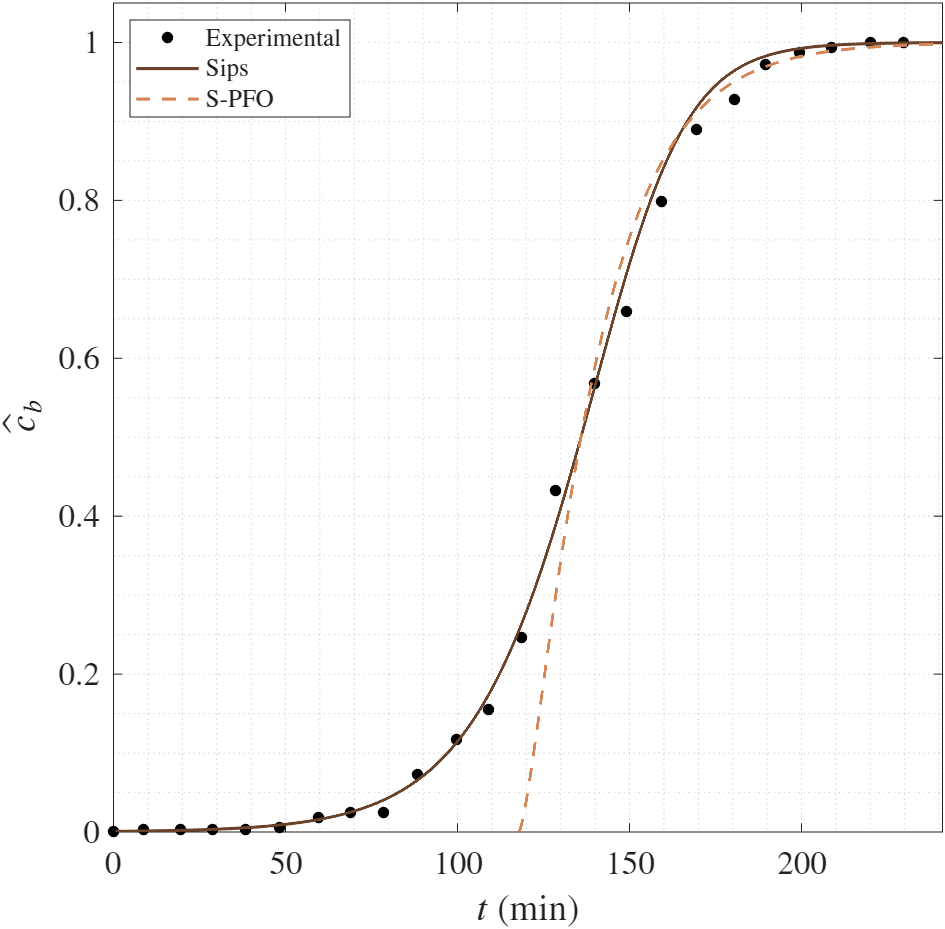}
    \put(90,13){(a)}
    \end{overpic}
    \begin{overpic}[width=.45\textwidth]{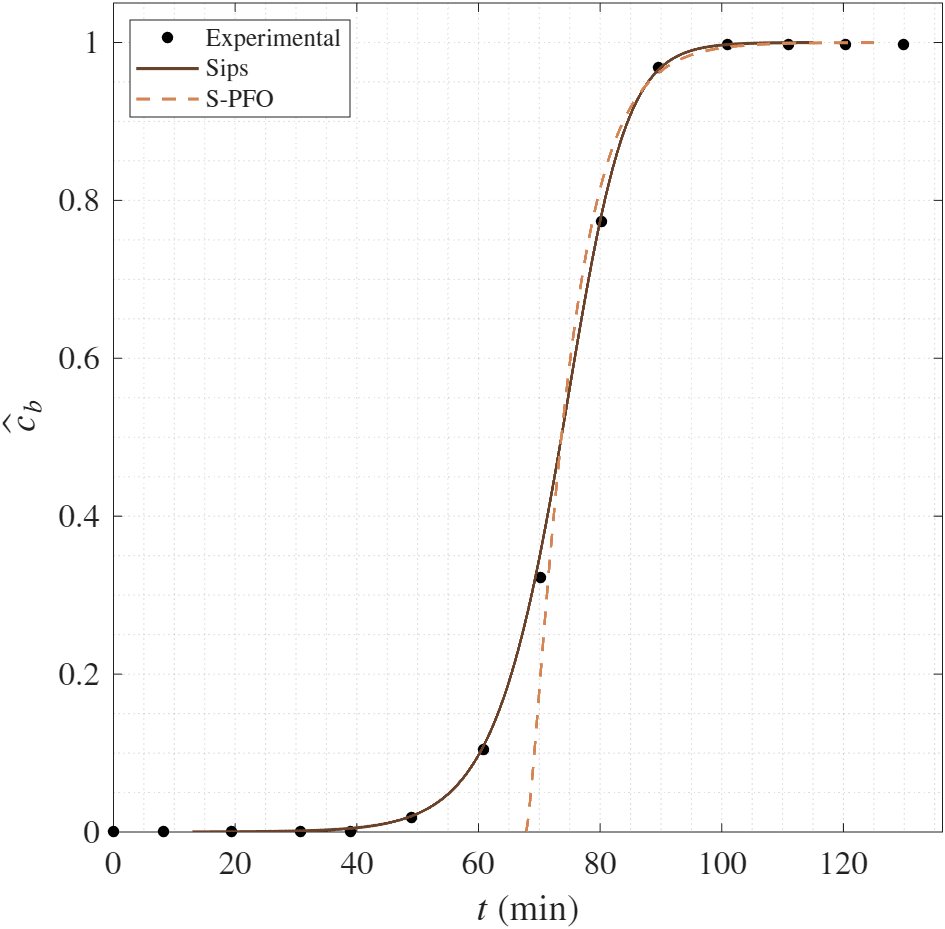}
    \put(90,13){(b)}
    \end{overpic}
    \begin{overpic}[width=.46\textwidth]{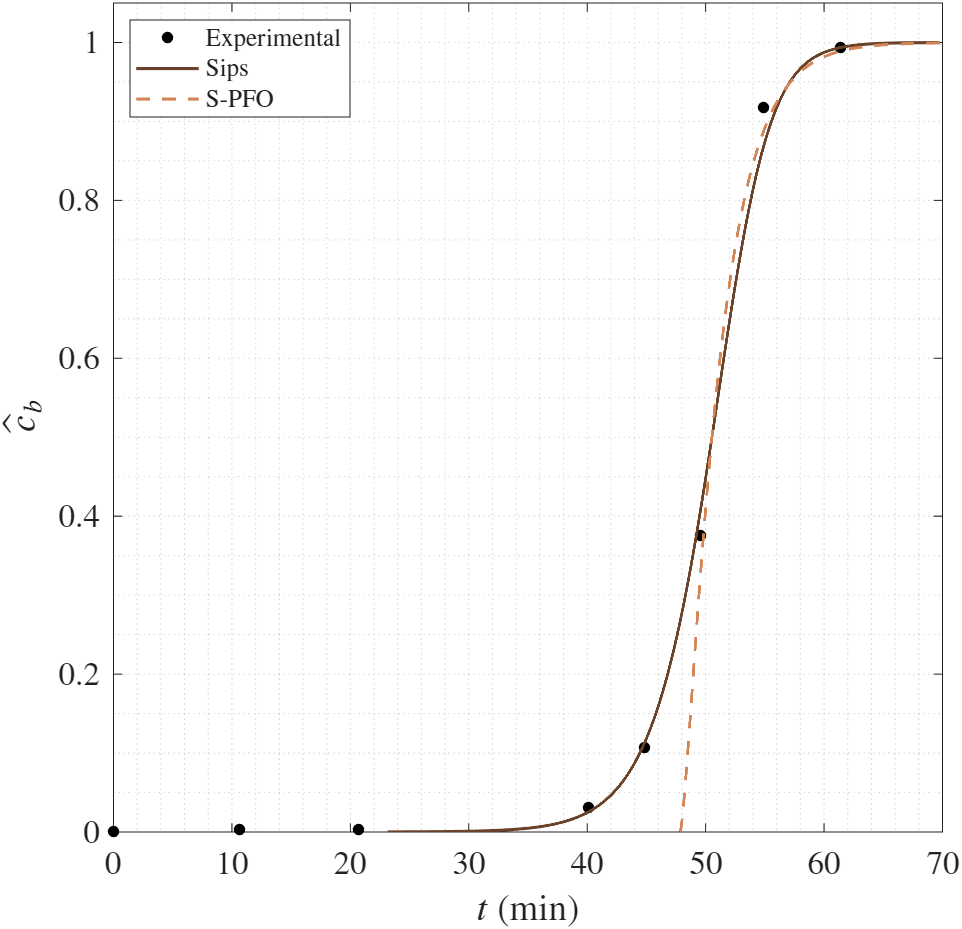}
    \put(90,13){(c)}
    \end{overpic}
    \begin{overpic}[width=.46\textwidth]{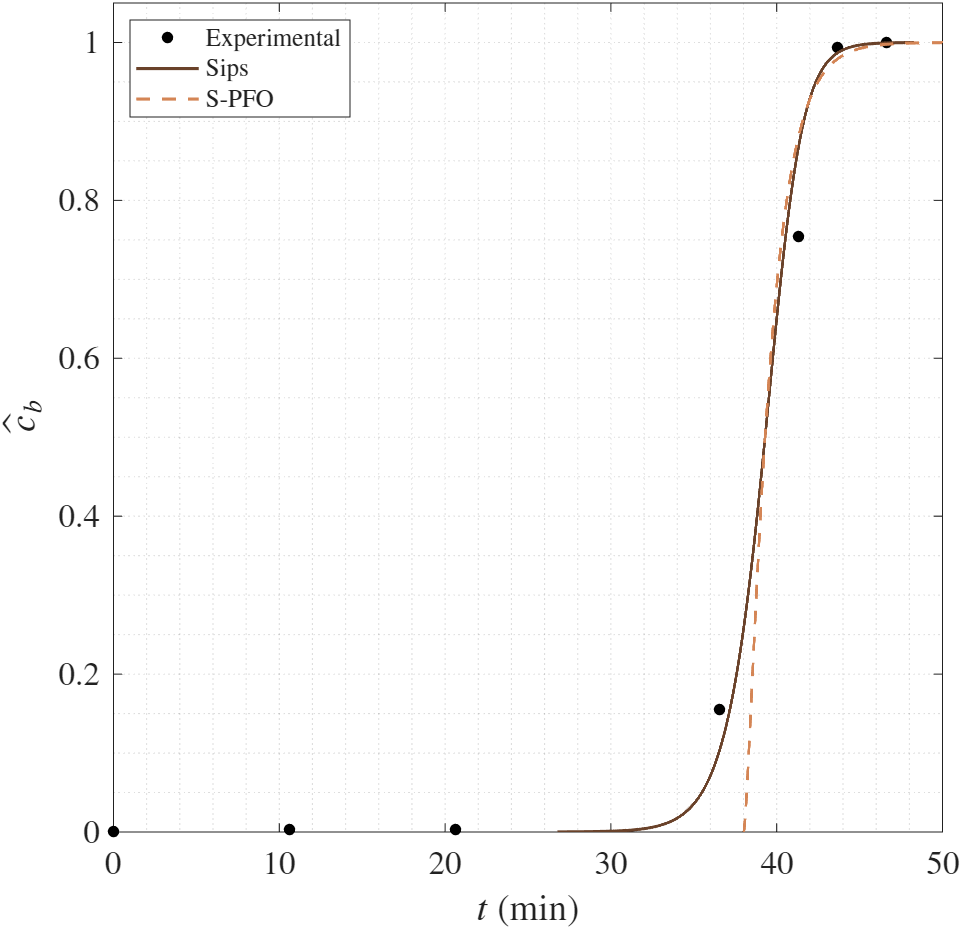}
    \put(90,13){(d)}
    \end{overpic}
    \caption{Breakthrough curves obtained for both Sips and S-PFO models with $(m,n)=(1,3)$ for different inlet concentrations, applied to experimental dataset 3 for the adsorption of mercury(II) onto dry activated sludge \cite{Sulaymon2014} (operating conditions in Table~\ref{Tab:data}). a) $c_{in}=0.13$ mol/m$^3$, b) $c_{in}=0.25$ mol/m$^3$, c) $c_{in}=0.37$ mol/m$^3$, d) $c_{in}=0.50$ mol/m$^3$.}
    \label{fig:HgSulaymon}
\end{figure}

The fitting of Sips and S-PFO model to breakthrough experimental datasets 1--3 (see Table~\ref{Tab:data}) for different inlet concentrations is shown in Figures~\ref{fig:TolRB3}, \ref{fig:SulaymonCEJ} and \ref{fig:HgSulaymon}. The fitted values of the adjustable parameters, namely $k_a$ for the Sips model and $k_{LDF}$ for the S-PFO model, together with the corresponding goodness-of-fit indicators, SSE and $R^2$, are reported in Table~\ref{tab:Fitted_result}. To provide more interpretable and directly comparable performance metrics, the goodness-of-fit indicators reported here are computed from the errors between the theoretical and experimental dimensionless breakthrough concentrations, $\hat{c}_b$. Note that the SSE actually minimised during the linear fitting procedure (and the corresponding $R^2$ values) differs from those reported here because its magnitude depends on the scale of the functions $\Psi_{mn}$ and $\Phi_\alpha$. The corresponding SSE and $R^2$ values obtained from the linear regression are reported in Table~\ref{tab:app_GoF_line}. Although these values differ from those listed in Table~\ref{tab:Fitted_result}, they exhibit the same increasing and decreasing trends across the different inlet concentrations and models.

\begin{table}[H]
\centering
\caption{Value of the fitted parameters using Sips and S-PFO models for different inlet concentrations for the three datasets presented in Table~\ref{Tab:data}. The sum of squared errors (SSE) and the coefficient of determination ($R^2$) obtained from the linear fitting of $\Psi_{mn}$ (Sips model) and $\Phi_\alpha$ (S-PFO model) are presented (see \ref{sec:appfittingprocedure}). The value of $q_e^{exp}$ is calculated using Eq.~\eqref{qeint}.}
\label{tab:Fitted_result}

\begin{tabular}{lcccc}
\multicolumn{5}{c}{Dataset 1 ($m=1$, $n=1$)} \\
\hline
$c_{\mathrm{in}}$ (mol m$^{-3}$)
& 0.0044 & 0.0104 & 0.0143 & 0.0308 \\
$q_e^{exp}$ (mol kg$^{-1}$)
& 2.6059 & 3.4187 & 3.3629 & 3.8161 \\
$k_a$ (m$^{3}$ mol$^{-1}$ s$^{-1}$)
& $1.12\times10^{-1}$ & $8.64\times10^{-2}$
& $1.15\times10^{-1}$ & $1.07\times10^{-1}$ \\
SSE & 0.0335 & 0.0239 & 0.0044 & 0.0179 \\
$R^2$ & 0.9991 & 0.9968 & 0.9993 & 0.9958 \\
$k_{\mathrm{LDF}}$ (s$^{-1}$)
& $6.05\times10^{-4}$ & $7.10\times10^{-4}$
& $1.09\times10^{-3}$ & $1.53\times10^{-3}$ \\
SSE & 0.1740 & 0.1212 & 0.0631 & 0.0707 \\
$R^2$ & 0.9951 & 0.9837 & 0.9899 & 0.9835 \\
\hline
\end{tabular}

\vspace{10pt}

\begin{tabular}{lccc}
\multicolumn{4}{c}{Dataset 2 ($m=1$, $n=2$)} \\
\hline

$c_{\mathrm{in}}$ (mol m$^{-3}$)
& 0.3934 & 0.7868 & 1.1802 \\
$q_e^{exp}$ (mol kg$^{-1}$)
& 0.0866 & 0.0997 & 0.0890 \\
$k_a$ (m$^{3}$ kg mol$^{-2}$ s$^{-1}$)
& $9.60\times10^{-3}$ & $7.50\times10^{-3}$ & $8.57\times10^{-3}$ \\
SSE
& 0.0192 & 0.0141 & 0.0431 \\
$R^2$
& 0.9932 & 0.9964 & 0.9833 \\
$k_{\mathrm{LDF}}$ (s$^{-1}$)
& $3.20\times10^{-4}$ & $3.70\times10^{-4}$ & $6.36\times10^{-4}$ \\
SSE
& 0.0625 & 0.0413 & 0.1458 \\
$R^2$
& 0.9780 & 0.9894 & 0.9434 \\
\hline
\end{tabular}

\vspace{10pt}

\begin{tabular}{lcccc}

\multicolumn{5}{c}{Dataset 3 ($m=1$, $n=3$)}
 
\\
\hline
$c_{\mathrm{in}}$ (mol m$^{-3}$)
 & 0.1246 & 0.2493 & 0.3739 & 0.4985 
\\
$q_e^{exp}$ (mol kg$^{-1}$)
 & 0.0184 & 0.0200 & 0.0204 & 0.0208 \\
$k_a$ (m$^{3}$ kg$^{2}$ mol$^{-3}$ s$^{-1}$)
& $1.03\times10^{2}$ & $9.17\times10^{1}$ & $1.07\times10^{2}$ & $1.46\times10^{2}$
\\
SSE
& 0.0089 & 0.0013 & 0.0037 & 0.0149
\\
$R^2$
 & 0.9979 & 0.9996 & 0.9984 & 0.9945
\\
$k_{\mathrm{LDF}}$ (s$^{-1}$)
 & $1.27\times10^{-3}$ & $3.87\times10^{-3}$ & $8.32\times10^{-3}$ & $1.72\times10^{-2}$
\\
SSE
 & 0.1283 & 0.0285 & 0.0161 & 0.0398
\\
$R^2$
 & 0.9691 & 0.9902 & 0.9930 & 0.9853
\\

\hline

\end{tabular}

\end{table}

The most immediate observation from Figures~\ref{fig:TolRB3}--\ref{fig:HgSulaymon} is that the Sips model provides a better agreement with the experimental data than the S-PFO model. In most cases, the largest discrepancies in the S-PFO predictions occur around the breakthrough time, as the resulting curves either lack symmetry (particularly for $(m,n)=(1,1)$) or fail to reproduce the smooth initial tail of the experimental breakthrough curves. This is further supported by the consistently higher SSE and lower $R^2$ values obtained for the S-PFO model across all cases (see Table~\ref{tab:Fitted_result}).

\begin{figure}[H]
    \centering
    \begin{overpic}[width=0.49\linewidth]{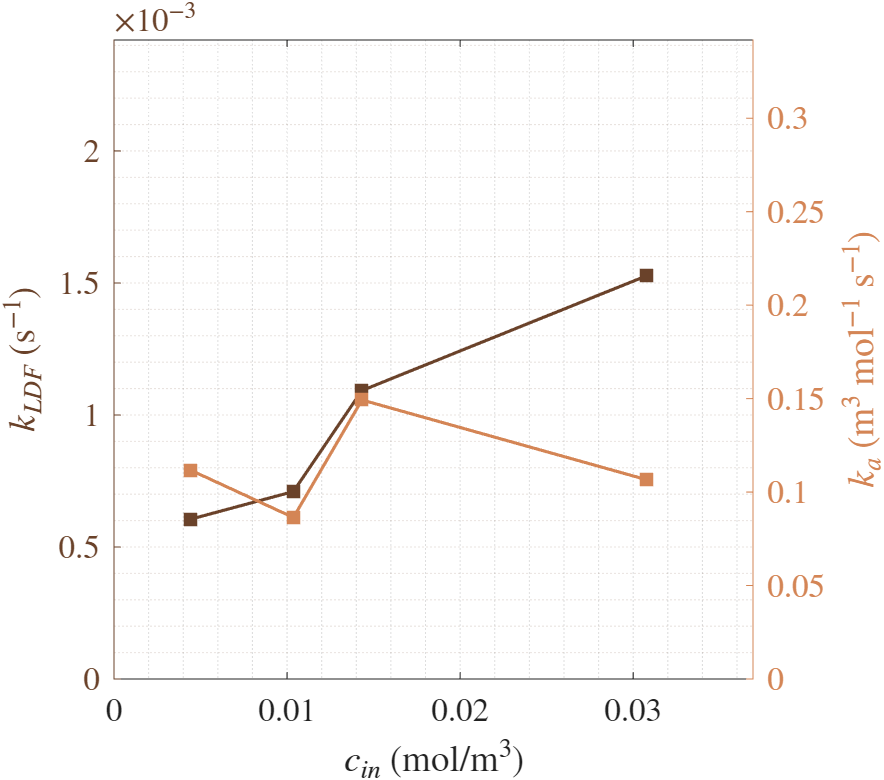}
    \put(92,15){}
    \end{overpic}%
    \hfill
    \begin{overpic}[width=0.49\linewidth]{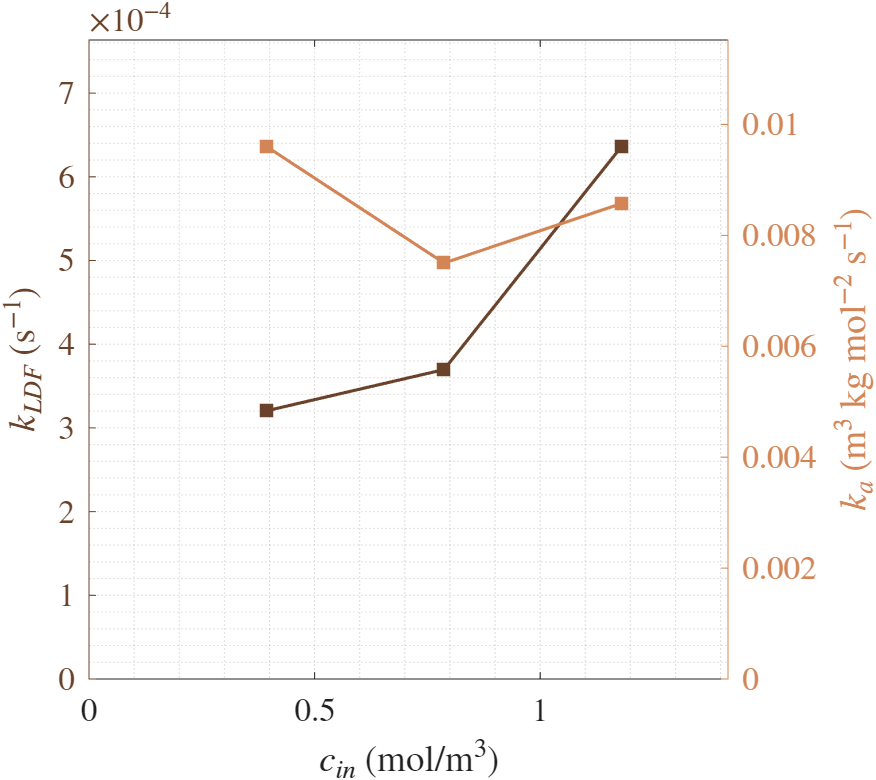}
    \put(92,15){}
    \end{overpic}%
    \caption{Concentration dependence of the S-PFO model parameter $k_{LDF}$ (left vertical axis) and the Sips model parameter $k_a$ (right vertical axis) for datasets 1 \cite{Myers23} (left) and 2 \cite{Sulaymon2009} (right).} 
    \label{fig:compare-ka-kLDF}
\end{figure}

However, for dataset 2 and for the lowest inlet concentration of dataset 1 (Figures~\ref{fig:TolRB3}a and \ref{fig:SulaymonCEJ}), both models provide a similarly good fit to the experimental data. For example, the difference in $R^2$ between the Sips and S-PFO models is only 0.4\% for dataset 1 with $c_{\text{in}}=0.0044$ mol/m$^3$, and 0.7\% for dataset 2 with $c_{\text{in}}=0.7868$ mol/m$^3$. Although these results may suggest that both models are equally capable of describing the experimental observations, only the Sips model remains physically consistent. Evidence of this can be found in Table~\ref{tab:Fitted_result}, where the fitted adsorption coefficient, $k_a$, remains approximately constant across the different inlet concentrations for all datasets. 
For example, across the different inlet concentrations, the fitted value of $k_a$ varies by at most a factor of 1.3, 1.1, and 1.6 for datasets 1, 2, and 3, respectively, with only minor fluctuations about its mean value.
In contrast, the fitted coefficient $k_{LDF}$ monotonically increases with increasing inlet concentration. In contrast, the fitted value of $k_{LDF}$ increases by factors of 2.5, 2.0, and 13.5 for datasets 1, 2, and 3, respectively, as the inlet concentration increases by factors of 7, 3, and 4.
This dependence is illustrated for datasets 1 and 2 in Figure~\ref{fig:compare-ka-kLDF}.

This is a fundamental limitation of the S-PFO model that is often overlooked in the literature. The dependence of the fitted adsorption coefficient on the inlet concentration indicates that the governing sink term is either missing relevant variables or does not have the correct functional form. By definition, adsorption and desorption rate coefficients should be independent of concentration, as the effect of concentration should already be accounted for by the remaining variables in the sink term. As shown in Table~\ref{tab:Fitted_result}, this requirement is satisfied by the Sips model. Consequently, fitting all breakthrough curves using a single value of $k_a$ for the different values of $c_{\text{in}}$ would still provide a reasonably good description of the experimental data. This approach has also been successfully applied in other adsorption systems (see Figure 3 in Myers et al. \cite{TwoAnalyRev}).

\section{Conclusions}\label{sec:6}

In this paper we have provided mathematical and experimental evidence demonstrating the fundamental inconsistency of the S-PFO model, specifically when coupled with the Sips isotherm (S-PFO), with respect to the physical interpretation of its kinetic parameters. While our analysis showed that the S-PFO model can sometimes yield fits to experimental data that appear comparable to the consistent Sips model—yielding $R^2$ differences as small as 0.4\% for Toluene adsorption and 0.7\% for Copper(II) adsorption—this apparent success is purely mathematical curve-fitting.  The structural flaw of the S-PFO model is exposed when examining the consistency of the extracted parameters. In the physically consistent Sips model, the fitted adsorption coefficient ($k_a$) remains remarkably stable across varying inlet concentrations, varying by a factor of at most 1.3, 1.1, and 1.6 for the datasets considered in this study. In stark contrast, the fitted mass transfer coefficient ($k_{LDF}$) in the S-PFO model is highly dependent on the inlet concentration, proving that it absorbs the underlying structural inconsistencies rather than representing a true physical parameter.

Despite these fundamental physical flaws, we were nonetheless able to solve the S-PFO model analytically using a travelling wave approximation, as done in previous studies with the Langmuir and Sips models \cite{Aguareles2022,SensiTW,Myers23}. This mathematical derivation further proves that the S-PFO model is structurally distinct with respect to the dynamic description, predicting an abrupt breakthrough rather than the continuous, smooth increase at the outlet. These differences are not critical by themselves, as data sometimes exhibits similar behaviours. However, the dependence on the the inlet concentration of the rate $k_{LDF}$ unequivocally demonstrates that a good fit to a single breakthrough curve is insufficient to validate the S-PFO as a kinetic model. 

We hope that this work encourages the adoption of physically consistent mathematical models and discourages the use of inappropriate formulations for describing adsorption kinetics.

\section*{Declaration of competing interest}
The authors declare that they have no known competing financial interests or personal relationships that could have appeared to influence the work reported in this paper.

\section*{Author contributions}
M. Calvo-Schwarzwalder: Conceptualization, Formal analysis, Investigation, Supervision, Writing - original draft. 
A. Valverde: Data curation, Investigation, Software, Validation, Writing - original draft.
A. Cuesta-López: Formal analysis, Investigation, Writing - original draft.
A. Cabrera-Codony: Data curation, Methodology, Resources, Writing - review \& editing.
U. Thorat: Software, Validation, Writing - original draft.
T.G. Myers: Conceptualization, Funding acquisition, Project administration, Writing - review \& editing.

\section*{Acknowledgements}
This publication is part of the research projects Minerva {(PID2023-146332OB-C21)} financed by {MCIN/AEI/ 10.13039/501100011033/}, by “ERDF A way of making Europe” and by “European Union NextGenerationEU/PRTR”. 
A. Cabrera-Codony acknowledges financial support from the Catalan government (2021-SGR-01352).
This work is supported by the Spanish State Research Agency, through the Severo Ochoa and Maria de Maeztu Program for Centres and Units of Excellence in R\&D (CEX2020-001084-M). 
A. Valverde and M. Calvo-Schwarzwalder are Serra-Hunter fellows from the Serra-Hunter Programme of the Generalitat de Catalunya. 
T. G. Myers and U. Thorat thank CERCA Programme/Generalitat de Catalunya for institutional support.
Open access funding was provided thanks to the CRUE-CSIC agreement with Elsevier. 

\appendix

\section{Existence of an additional root of $g_\alpha$ for $\alpha>1$}\label{app:v alpha}
will find a value $\hat v_\alpha$ such that if $\hat v\in(1,\hat v_\alpha)$, then the solution of Eq.~\eqref{TW:odePFO} does not satisfy the far-field conditions of Eq.~\eqref{app:proof:conditions}. That is, there exists a value $\mathcal{C}^*\in(0,1)$ such that $g_{\alpha}(\mathcal{C}^*) = 0$.

Since $\alpha > 1$, evaluating $h_\alpha$ at the boundaries gives
\begin{equation} 
    h_\alpha(0) = 1-\hat v<0,\qquad h_\alpha(1)=0. 
\end{equation}
Again, the key is to observe the behaviour of the derivative of $h_\alpha$, defined in Eq.~\eqref{TW:diff_h}. 
For a root $\mathcal{C}^*$ to exist in $(0, 1)$, $h(\mathcal{C})$ must become positive before returning to zero at $\mathcal{C}=1$. This is guaranteed if $\ud h/\ud\mathcal{C}$ is negative as $\mathcal{C}\to1^-$, meaning the peak occurs before the endpoint:
\begin{equation}\label{TW:cond v alpha}
    \nd{h_\alpha}{\mathcal{C}}\bigg|_{\mathcal{C}=1}<0\quad\Longleftrightarrow\quad\hat v<\frac{\alpha}{\alpha-1}=:\hat v_\alpha.
\end{equation}
In summary, we have proved that for $\alpha > 1$, a third zero $\mathcal{C}^* \in (0, 1)$ exists if and only if $\hat v \in(1, \hat v_\alpha)$. In this situation, the travelling wave solution does not exist. In the limit as $\alpha\to\infty$, this reduces to $\hat v_\alpha\to1$, which is always satisfied. Conversely, as $\alpha\to1^+$ we have $\hat v_\alpha\to\infty$ and thus the existence of a travelling wave requires larger velocities, which is unphysical as the flow is typically assumed to be slow. 

\section{Convergence of $\Phi_\alpha(0^+)$ for $\alpha<1$}\label{appB}

Consider the integral defined by 
\begin{equation}\label{appB:integral}
\Phi_{\alpha}(0^+) = \int_{1/2}^{0} \frac{1}{g_{\alpha}(s)} \, \ud s,
\end{equation}
where the function $g_{\alpha}(s)$ is defined as
\begin{equation}
g_{\alpha}(s) = \frac{\hat{v}s^{\alpha} - (s^{\alpha} + \hat{v} - 1)s}{\hat{v}(s^{\alpha} + \hat{v} - 1)}
\end{equation}
with $\hat v>1$. If $\alpha<1$, then $\Phi_{\alpha}(0^+)<\infty$. 
To show prove this result we must look at the asymptotic behaviour of $1/g_\alpha$ as $s \to 0^+$, since $g_\alpha(0)=0$.  The function $1/g_\alpha$ behaves asymptotically as
\begin{equation}
\frac{1}{g_{\alpha}(s)} \sim \frac{\hat{v}(\hat{v}-1)}{\hat{v}s^{\alpha}} = \frac{\hat{v}-1}{s^{\alpha}}.
\end{equation}
Since $\alpha<1$, we have
\begin{equation}
    \int_{0}^{1/2} \frac{1}{s^{\alpha}} \, ds <\infty,
\end{equation}
and therefore the integral in Eq.~\eqref{appB:integral} also converges. That is, given $\gamma>0$ there exists a finite value $\eta^*_\alpha>0$ such that
\begin{equation}
-\Phi_{\alpha}(0^+)=\gamma\eta^*_\alpha<+\infty.
\end{equation}

\section{Fitting procedure of experimental breakthrough curves} \label{sec:appfittingprocedure}

We first perform a transformation of the breakthrough data of the form 
\begin{equation}
\Phi^{exp}_{\alpha,i}:=\Phi_\alpha(c_{b,i}^{exp}),\qquad\Psi^{exp}_{mn,i}:=\Psi_{mn}(c_{b,i}^{exp}). 
\end{equation}
The fitting procedure then reduces to a simple linear fit and we obtain the optimal value $b^*$ via the least-square method, 
\begin{equation}
    \min_{b>0}\left\{\sum_{i=1}^{N} \left(Y_i^{\text{exp}}-bX^{exp}_i\right)^2\right\}
\implies
    {b}^* = \frac{\sum_{i=1}^{N} (X_i^{\text{exp}} - \bar{X})(Y_i^{\text{exp}} - \bar{Y})}{\sum_{i=1}^{N} (X_i^{\text{exp}} - \bar{X})^2}\,, \label{eq:fitting}
\end{equation}
where $X_i^{exp}=t_i^{exp}-t_{h}^{exp}$, $Y^{exp}_i=\Phi^{exp}_{\alpha,i}$ and $b=k_{LDF}(1+\kappa^{1/n})$ for the S-PFO model, $Y^{exp}_i=\Psi_{mn,i}^{exp}$ and $b=k_a(1+\k^{1/n})q_\text{max}^{n-1}c_\text{in}^m$ for the original Sips formulation, and the quantity $\bar{A}$ represents the mean of the data $A_1,\ldots,A_N$. Lastly, the goodness measures for this fit are then
\begin{equation}
    SSE= \sum_{i=1}^{N} \left(Y_i^{\text{exp}}-b^*X^{exp}_i\right)^2\, ,\qquad R^2=1-\frac{SSE}{\sum_{i=1}^{N} (Y_i^{\text{exp}} -\bar Y)^2}.
\end{equation}

The goodness-of-fit indicators for the linear regression described above are reported in Table~\ref{tab:app_GoF_line}. Note that the order of magnitude of the sum of squared errors (SSE) depends on the magnitude of the functions $\Psi_{mn}$ and $\Phi_\alpha$. The coefficient of determination, $R^2$, also differs from the values reported in Table~\ref{tab:Fitted_result}; however, it exhibits the same increasing and decreasing trends across the different inlet concentrations and models.

\begin{table}[H]
\centering
\caption{Goodness of the fit indicators obtained from the original linear fitting of $\Psi_{mn}$ (Sips model) and $\Phi_\alpha$ (S-PFO model) for the three datasets presented in Table~\ref{Tab:data} following the procedure in Section~\ref{sec:appfittingprocedure} }
\label{tab:app_GoF_line}

\begin{tabular}{lccccc}
\multicolumn{6}{c}{Dataset 1 ($m=1$, $n=1$)} \\
\hline
\multirow{2}{3em}{Sips} & 
SSE & 4.374 & 2.516 & 0.152 & 3.819 \\
 & $R^2$ & 0.9870 & 0.9534 & 0.9942 & 0.9063 \\
\hline
\multirow{2}{3em}{S-PFO} & 
SSE & 38.29 & 7.878 & 1.040 & 8.603 \\
 & $R^2$ & 0.9206 & 0.7931 & 0.8798 & 0.4842 \\
\hline
\end{tabular}

\vspace{10pt}

\begin{tabular}{lcccc}
\multicolumn{5}{c}{Dataset 2 ($m=1$, $n=2$)} \\
\hline
\multirow{2}{3em}{Sips} & 
SSE
& 2.188 & 6.551 & 3.179 \\
 & $R^2$
& 0.9916 & 0.9828 & 0.9851 \\
\hline
\multirow{2}{3em}{S-PFO} & 
SSE
& 5.789 & 3.773 & 5.526 \\
 & $R^2$
& 0.9492 & 0.9815 & 0.9523 \\
\hline
\end{tabular}

\vspace{10pt}

\begin{tabular}{lccccc}
\multicolumn{6}{c}{Dataset 3 ($m=1$, $n=3$)}
 
\\
\hline
\multirow{2}{3em}{Sips} & 
SSE
& 130.3$\times10^{5}$ & 0.57$\times10^{5}$ & 1.03$\times10^{5}$ & 1.74$\times10^{5}$
\\
 & $R^2$
 & 0.9888 & 0.9990 & 0.9932 & 0.9446
\\
\hline
\multirow{2}{3em}{S-PFO} & 
SSE
 & 240.7 & 235.4 & 160.7 & 242.1
\\
 & $R^2$
 & 0.9453 & 0.8954 & 0.8699 & 0.7176
\\

\hline

\end{tabular}

\end{table}

\begin{figure}[H]
    \centering
    \begin{overpic}[width=.41\textwidth]{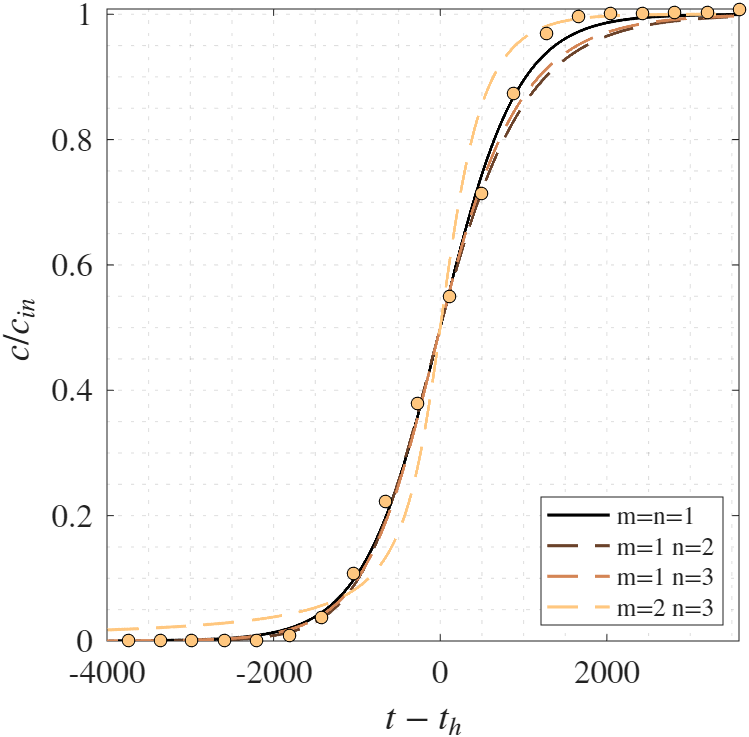}
    \put(15,92){a)}
    \end{overpic}
    \begin{overpic}[width=.41\textwidth]{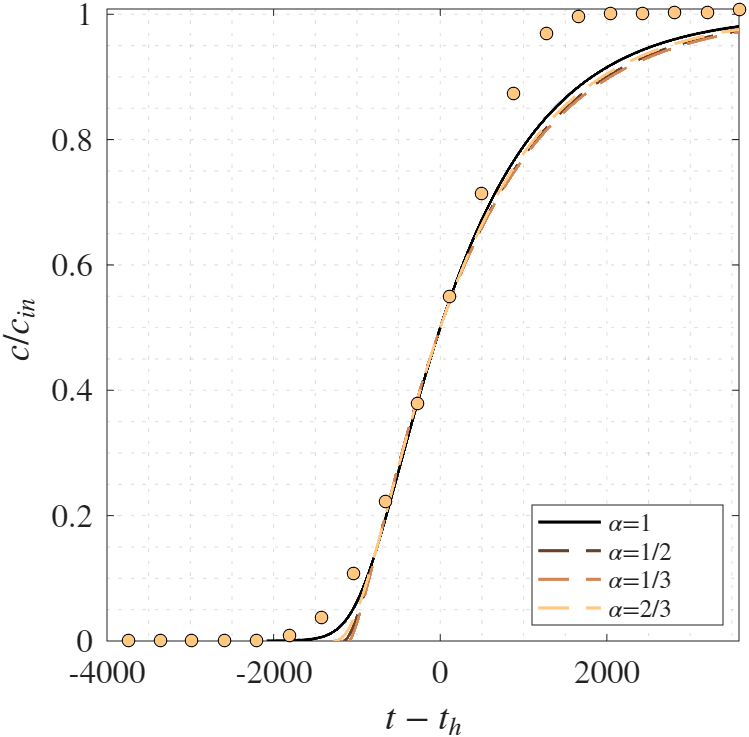}
    \put(15,92){b)}
    \end{overpic}
    \begin{overpic}[width=.41\textwidth]{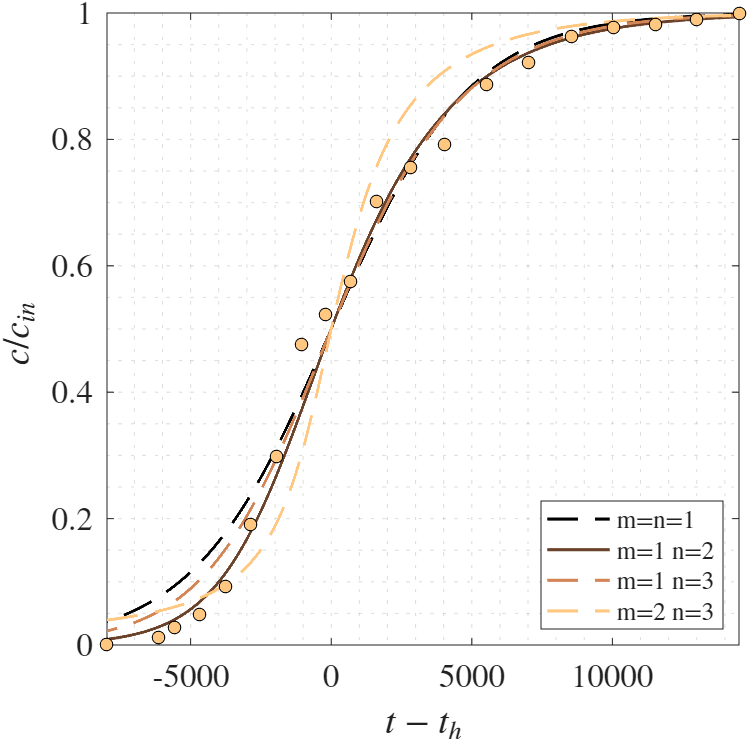}
    \put(15,92){c)}
    \end{overpic}
    \begin{overpic}[width=.41\textwidth]{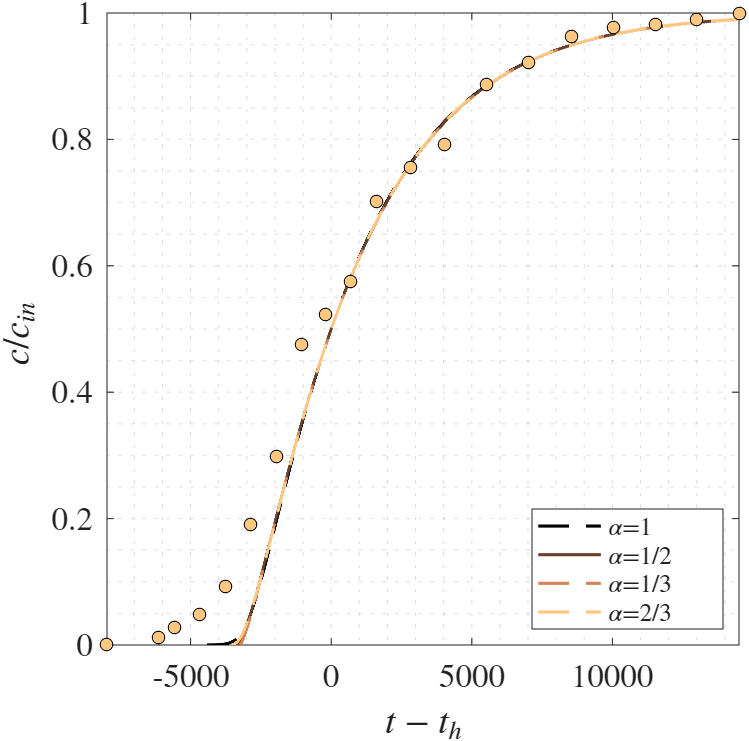}
    \put(15,92){d)}
    \end{overpic}
    \begin{overpic}[width=.41\textwidth]{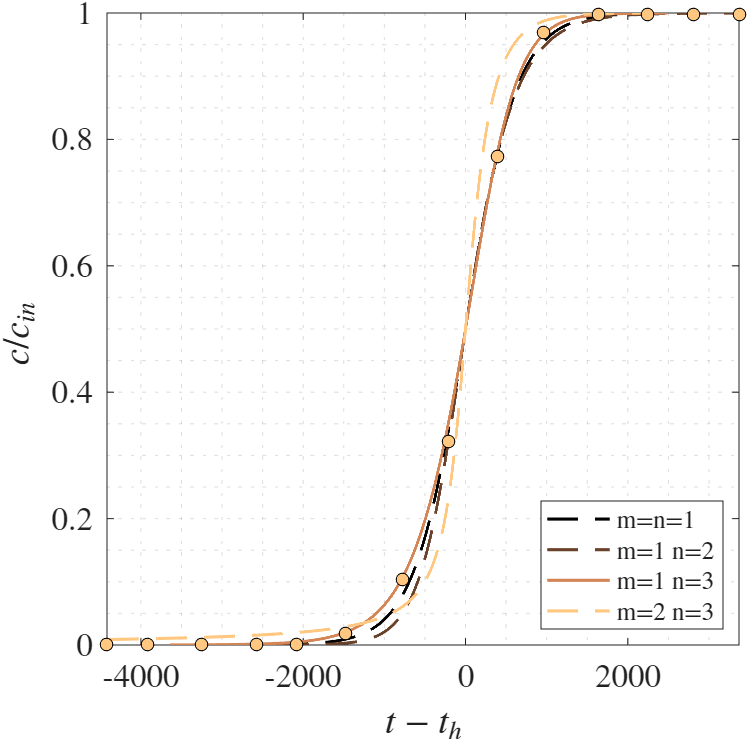}
    \put(15,92){e)}
    \end{overpic}
    \begin{overpic}[width=.41\textwidth]{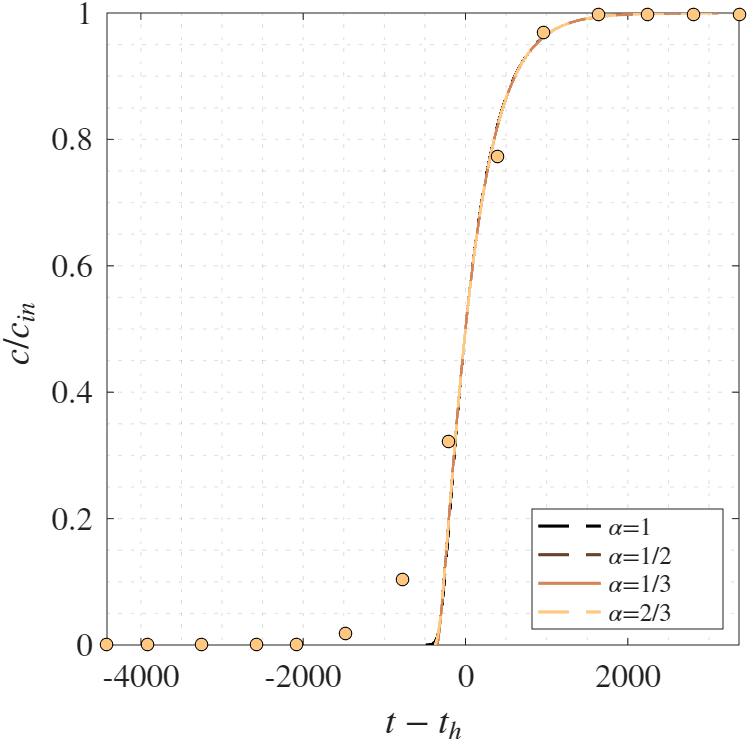}
    \put(15,92){f)}
    \end{overpic}
    \caption{Breakthrough curves obtained for both Sips and S-PFO models with different $(m,n)$ combinations for different datasets (see Table~\ref{Tab:data}). Dataset 1 $c_{in}=0.0104$ mol/m$^3$ a) Sips, b) S-PFO. Dataset 2 $c_{in}=0.39$ mol/m$^3$ a) Sips, b) S-PFO. Dataset 3 $c_{in}=0.25$ mol/m$^3$ a) Sips, b) S-PFO.}
    \label{fig:appBCmn}
\end{figure}

In Figure~\ref{fig:appBCmn} the fitting to experimental breakthrough data using different combinations $(m,n)$ is shown for datasets 1 to 3 (see Table~\ref{Tab:data}).

\section{Isotherms study} \label{sec:appiso}

In column studies, the isotherm values are obtained by fitting the isotherm equation to the available data points $(c_\text{in},q_\text{e}^{exp})$ obtained after performing the experiment repeatedly with different inlet concentrations while fixing the rest of operational conditions.

For a given experiment, the measurements at the outlet are of the form 
\begin{equation}
    D=\{(t_i^{exp},c_{b,i}^{exp})\, |\, i=1\ldots N\}\cup\{(t_h^{exp},c_\text{in}/2)\}\, ,
\end{equation}
where the half time is either measured or computed via an interpolation of the breakthrough data. In addition, we compute the final adsorbed quantity $q_\text{e}^{exp}$ via numerical integration of the breakthrough data until the experiment finishes $t=t_f$,
\begin{align}\begin{split}
    q_\text{e}^{exp}&=\frac{Q_\text{in}}{m_{ad}}\int_0^{t_f}c_\text{in}-c_b^{exp}\ud t\\
    &\approx \frac{Q_\text{in}}{m_{ad}}\left[c_\text{in}t_f - \sum_{i=1}^{n-1} \left( \frac{c^{exp}_{b,i} + c^{exp}_{b,i+1}}{2} \right) \Delta t_i \right], \label{qeint}
\end{split}\end{align}
where $Q_\text{in}$ and $m_{ad}$ are the volumetric flow rate and initial mass of adsorbent in the column and $\Delta t_i = t^{exp}_{i+1} - t^{exp}_i$.
Once the pairs $(c_\text{in},q_\text{e}^{exp})$ are available, we can obtain $\k$ and $q_\text{max}$ by adjusting Eq.~\eqref{eq:isosips} against these data points.

Although this is the standard approach in column (dynamic) studies, several authors perform dedicated equilibrium experiments in batch (static) systems to determine the adsorption isotherm. In these experiments, the adsorbent is placed in a well-mixed vessel and contacted with solutions of varying concentrations. Once equilibrium is reached, the equilibrium concentration in the liquid phase ($c_e^{exp}$) and the corresponding adsorbed amount ($q_e^{exp}$) are measured. It should be noted that isotherms derived from dynamic experiments, expressed as $(c_{in}, q_e^{exp})$, and those obtained from static experiments, expressed as $(c_e^{exp}, q_e^{exp})$, do not necessarily coincide. In fact, batch isotherms typically yield lower values of $q_{\text{max}}$ \cite{Auton2024,Auton2025}. Nevertheless, according to equilibrium thermodynamics, the equilibrium constant should remain unchanged. Consequently, the parameter $q_e^{exp}/q_{\text{max}}$ is expected to be applicable to both systems. Therefore, for the experimental datasets considered in this section, whenever well-defined isotherms obtained from dedicated equilibrium experiments were available—whether from dynamic or static setups—they were used preferentially to determine the isotherm parameters. In cases where such data were unavailable, the isotherm was reconstructed from the available datasets using the approach described in Eq.~\eqref{qeint}.

\begin{figure}[H]
\begin{center}
    \begin{overpic}[width=.31\textwidth]{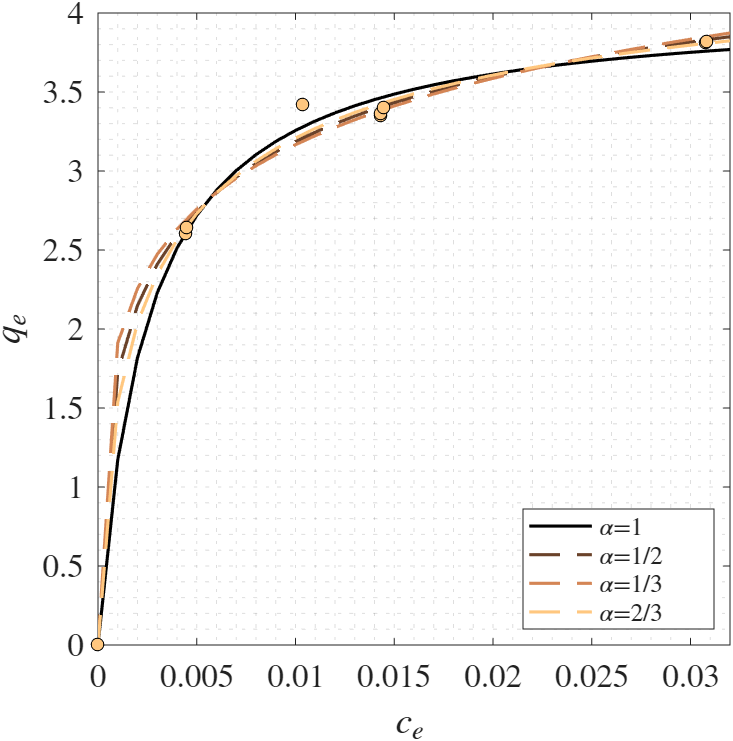}
    \end{overpic}%
    \hspace{1mm}
    \begin{overpic}[width=.32\textwidth]{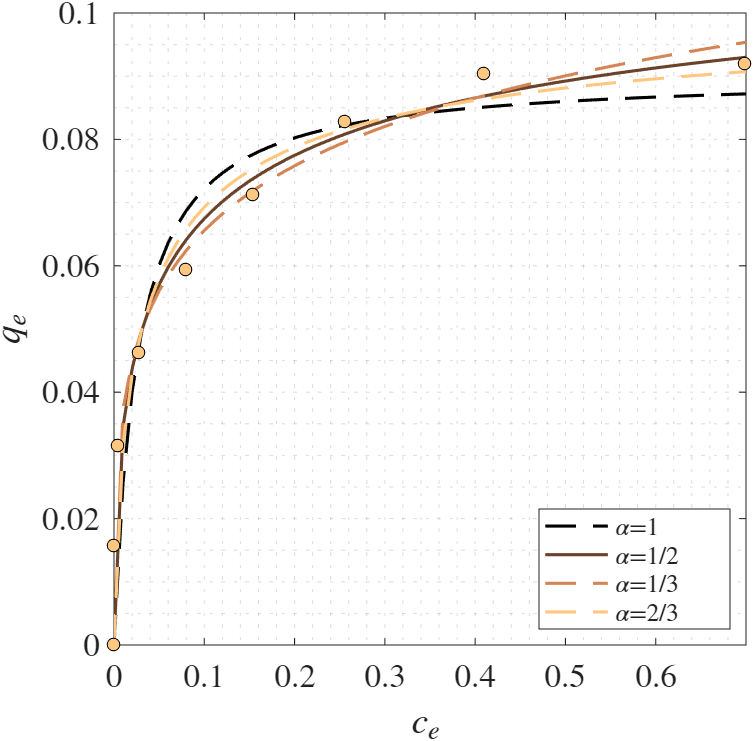}
    \end{overpic}%
    \hspace{1mm}
    \begin{overpic}[width=.33\textwidth]{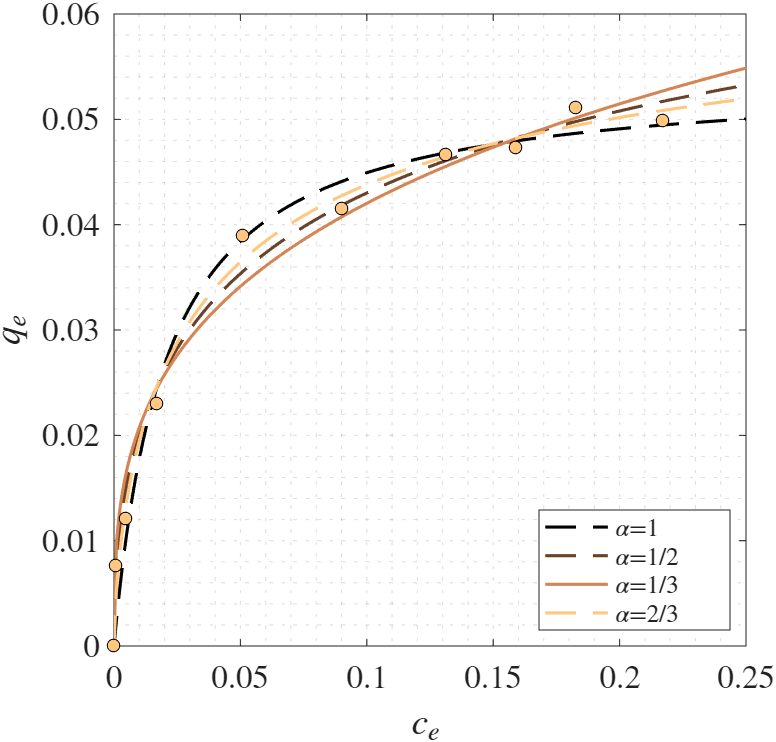}
    \end{overpic}
\end{center}
    
    \caption{Isotherm curves obtained with Eq.~\eqref{eq:isosips} for different values of $\alpha$, applied to equilibrium experimental data of toluene adsorption onto SAC by Myers et al. \cite{Myers23}, copper(II) adsorption onto GAC by Sulaymon et al. \cite{Sulaymon2009}, and mercury(II) adsorption onto DAS by Yousif et al. \cite{Yousif2013}.}
    \label{fig:isotherms}
\end{figure}

Figure~\ref{fig:isotherms} and Table~\ref{tab:appgofiso} present the results of fitting the isotherm expression given by Eq.~\eqref{eq:isosips}, for different values of $\alpha$, to the equilibrium adsorption data for toluene onto SAC reported by Myers et al. \cite{Myers23}, copper(II) onto GAC reported by Sulaymon et al. \cite{Sulaymon2009}, and mercury(II) onto DAS reported by Yousif et al. \cite{Yousif2013}, which includes the batch adsorption isotherm corresponding to the fixed-bed column experiments presented by Sulaymon et al. \cite{Sulaymon2014}. The isotherm parameters corresponding to the final $(m,n)$ orders selected for the breakthrough analysis are listed in Table~\ref{tab:isotherm_parameters}.

\begin{table}[H]
\centering
\caption{Goodness of the fit of the adjustment of  Eq.~\eqref{eq:isosips} to equilibrium experimental data for different values of $\alpha$.}
\label{tab:appgofiso}

\begin{tabular}{lccccc}
\hline
\textbf{Author} & \textbf{Coefficient} & $\mathbf{\alpha=1}$ & $\mathbf{\alpha=1/2}$ & $\mathbf{\alpha=1/3}$ & $\mathbf{\alpha=2/3}$ \\
\hline
\multirow{2}{10em}{Myers et al.~\cite{Myers23}} & SSE ($\times10^{-2}$) & 5.501 & 5.331 & 6.663 & 4.742 \\
& $R^2$ & 0.9951 & 0.9952 & 0.9940 & 0.9958 \\
\hline
\multirow{2}{10em}{Sulaymon et al.~\cite{Sulaymon2009}} & SSE ($\times10^{-4}$)
& 7.990 & 3.505 & 3.033 & 4.636 \\
& $R^2$
& 0.9099 & 0.9605 & 0.9658 & 0.9477 \\
\hline
\multirow{2}{10em}{Yousif et al.~\cite{Yousif2013}} & SSE ($\times10^{-5}$)
 & 5.798 & 3.087 & 5.249 & 2.942
\\
& $R^2$
 & 0.9828 & 0.9908 & 0.9844 & 0.9913
\\

\hline

\end{tabular}

\end{table}

\begin{table}[H]
\centering
\caption{Value of the equilibrium parameters obtained for different values of $\alpha$ adjusted by fitting Eq.~\eqref{eq:isosips} to equilibrium experimental data of toluene adsorption onto SAC by Myers et al. \cite{Myers23}, copper(II) adsorption onto GAC by Sulaymon et al. \cite{Sulaymon2009}, and mercury(II) adsorption onto DAS by Yousif et al. \cite{Yousif2013}.}
\label{tab:isotherm_parameters}

\begin{tabular}{lccc}
\hline
 & Dataset 1 & Dataset 2 & Dataset 3 \\
\hline
Author & Myers et al. \cite{Myers23} & Sulaymon et al. \cite{Sulaymon2009} & Yousif et al. \cite{Yousif2013} \\

Type & Column & Batch & Batch\\

$\alpha$
& 1 & 1/2 & 1/3 \\

$q_{\max}$ (mol/kg)
& 4.059 & 0.1208 & 0.3789 \\

$K_S$ (m$^{3\alpha}$/mol$^\alpha$)
& 406.09 & 4.0016 & 0.2688 \\

SSE (mol$^2$/kg$^2$)
& 0.055 & $3.50\times10^{-4}$ & $5.25\times10^{-5}$ \\

$R^2$
& 0.9951 & 0.9605 & 0.9844 \\

\hline
\end{tabular}

\end{table}


\end{document}